\documentclass[journal]{IEEEtran}
\usepackage[utf8]{inputenc} 
\usepackage[T1]{fontenc}
\usepackage{cite}

\usepackage[pdftex]{graphicx}
\DeclareGraphicsExtensions{.pdf}

\usepackage[cmex10]{amsmath}
\usepackage{amssymb}
\usepackage{mathrsfs}

\usepackage[noend]{algpseudocode}
\usepackage{algorithm}
\floatstyle{plaintop} \restylefloat{algorithm}

\usepackage{framed}
\usepackage{paralist}
\usepackage{booktabs}
\usepackage{array}
\newcolumntype{C}[1]{>{\centering\let\newline\\\arraybackslash\hspace{0pt}}m{#1}}
\usepackage{mdwmath}
\usepackage{eqparbox}

\usepackage[usenames,dvipsnames]{xcolor}
\usepackage[caption=false,font=normalsize,labelfont=sf,textfont=sf]{subfig}

\newcommand \red[1]         {{\color{red}#1}}

\newcommand{\hide}[1]{}
\newcommand{\PP}{\mathbf{P}}
\newcommand{\XX}{\mathcal{X}}
\newcommand{\YY}{\mathcal{Y}}

\newcommand{\SecSet}{\red{\Theta}}

\newcommand{\RR}{\mathbb{R}}
\newcommand{\Env}{\mathcal{M}}

\newcommand{\diag}{\mathrm{diag}}

\newcommand{\preimage}{\operatorname{PreIm}}

\newtheorem{thm}{Theorem}
\newtheorem{lem}{Lemma}
\newtheorem{prop}{Proposition}
\newtheorem{corr}{Corollary}
\newtheorem{sub-lem}{Sub-lemma}
\DeclareMathOperator\supp{supp}
\begin{document}
\title{Generalised Entropies and Metric-Invariant Optimal Countermeasures for Information Leakage under Symmetric Constraints}
\author{\IEEEauthorblockN{MHR.~Khouzani,~\IEEEmembership{Member,~IEEE,} and
        Pasquale~Malacaria}\\
    \thanks{Both authors are with the School of Electronic Engineering and 
        Computer Science at Queen Mary University of London, London, UK. 
        Their emails are: \texttt{arman.khouzani@qmul.ac.uk} and 
        \texttt{p.malacaria@qmul.ac.uk}. 
        This work was supported by EPSRC grant EP/K005820/1 titled 
        ``Games and Abstraction: The Science of Cyber Security''.
    }%
    \thanks{A conference version of this work appeared in
        29th IEEE Computer Security Foundations Symposium (CSF)
        \cite{khouzani2016relative}. 
}}

\markboth{ACCEPTED TO IEEE TRANSACTIONS ON INFORMATION THEORY}%
{Khouzani \MakeLowercase{\textit{et al.}}: Generalised Entropies and Metric-Invariant Optimal Countermeasures}

\maketitle

\begin{abstract}
	We introduce a novel generalization of entropy and conditional entropy from which most definitions from the literature can be derived as particular cases. Within this general framework, we investigate the problem of designing countermeasures for information leakage. In particular, we seek metric-invariant solutions,  i.e., they are robust against the choice of entropy for quantifying the leakage.
	The problem can be modelled as an information channel from the system to an adversary, and the countermeasures can be seen as modifying this channel in order to minimise the amount of information that the outputs reveal about the inputs. Our main result is to fully solve the problem under the highly symmetrical design constraint that the number of inputs that can produce the same output is capped.
	Our proof is constructive and the optimal channels and the minimum leakage are derived in closed form. 
	
\end{abstract}
\IEEEpeerreviewmaketitle

\section{Introduction}

In many computational systems, the system's behaviour
is affected by a confidential internal state and produces some publicly observable
behaviour.
This can be modelled as an information channel where the input of the channel is the confidential state of the system and the output is the externally observable behaviour, which can be observed not only by the intended recipient but also by some malicious agent.
The security goal here is to minimize the  leakage of confidential state to potentially adversarial observers.

As a simple example of this problem consider a government website processing tax returns.
Suppose the website takes less than 10 seconds to process a tax return and send an electronic acknowledgement for an individual who owes less than \$1000 in tax,  but it takes 20 seconds to process a tax return and send an electronic acknowledgement for an individual who owes more than \$1000 in tax. Then an eavesdropper that can only observe the time it takes to produce the electronic acknowledgement can learn confidential information about the user.
A solution to avoid this leak could be to make sure that the website always sends the acknowledgement after 20 seconds. Then an attacker cannot observe any behavioural difference and so no information about the internal state, i.e., the user tax status, is disclosed.
This countermeasure can be seen as a channel that maps both secrets (owing less or more than \$1000 in tax) to the same observable (20 seconds to generate the acknowledgement). The pre-image of this map on its produced observable includes both possible secrets.

In many contexts, the trivial countermeasure
of mapping all secrets to a unique observable may
be unsuitable or even infeasible. Consider for instance a password-checker system: at the bare minimum, the system should produce two
distinct observables (password match/mismatch) to preserve its defining
functionality.
There are also cases where mapping all secrets to the same observable (and so zero leakage)
may be possible but undesirable as it leads to an unacceptable
degradation in the utility of the system. Some prominent
examples include location privacy \cite{ardagna2007location,khoshgozaran2009private,theodorakopoulos2014prolonging}, in which if a mobile device reports the same location coordinates, then it may receive no connectivity or unfavourably poor location-based service. Another example is defences against web traffic fingerprinting \cite{cai2014systematic,juarez2016toward}, where generating the same observable involve lengthening inter-packet delays or generating dummy packets, both of which can have an undesirably large bandwidth or delay overhead.
Motivated by this observation and still allowing for analytical treatment, we consider a highly symmetric constraint: that is, we restrict the size of the subsets of the inputs (secrets) that can be mapped to the same output (observable). This will disallow all the inputs to be conflated with each other through the same output as the trivial solution.
Our problem statement is then: given this pre-image size constraint, can we design a channel of minimal leakage?
This problem is a stylised abstraction of the above mentioned real world scenarios, with surprising mathematical properties explored in this paper.

A challenging problem in designing leakage-minimal channels is that there are several
candidates for quantifying information leakage, e.g., Shannon
\cite{clark2005quantitative}, Min-Entropy \cite{smith2009foundations},
Bayesian \cite{mciver2010compositional}, $g$-leakage \cite{alvim2012measuring},
guesswork (guessing) entropy \cite{massey1994guessing},
R\'enyi family \cite{fehr2014conditional}, \textit{etc}.
This is rather problematic, as some of  these entropies have distinct operational interpretation that  rely on different modelling of the behaviour or the abilities of the adversary \cite{khouzani2018optimal}. 
Therefore, we add this desirable notion of robust optimality to our design problem: the  channel should stay optimal with respect to any ``reasonable'' choice of  leakage quantification.

Given such a strong requirement of robust optimality, there is no a priori reason that such a solution should  even exist.
Moreover, the few robustness results in the field of quantification of leakage have been hard to prove
(e.g. the proof of the Coriaceous Conjecture \cite{mciver2014abstract}).
This work contributes to both leakage guarantees and robustness in that
it investigates channels that are ``metric-invariantly'' optimal within the large class of our generalised entropies.

\paragraph*{\textbf{Road-map and Contributions}}
The focus of this work is foundational. In particular, the list of our contributions
is as follows:
First we introduce a general framework which includes all reasonable entropies and derived leakage notions.  Specifically, our generalised entropies satisfy the basic properties of symmetry, expansibility, and a form of concavity as made precise later in the paper.
We show for example that Sharma-Mittal entropies (themselves including R\'enyi entropies), Arimoto conditional entropies, Guesswork and other known measures from the literature are particular cases of our definition.

Second,
we formalize the problem of minimizing the information leakage given a prior
distribution of the secret and constraints on how many secrets can be mapped to a common output, where the information
leakage is quantified as the difference between the prior and posterior uncertainties of
an adversary for our generic entropy function (Section~\ref{sec:Model}). In Section~\ref{sec:Analysis},
we express and prove our main result (Theorem~\ref{Thm:main}), that is, we provide the lowest
achievable leakage across all (potentially probabilistic) channels in closed form.
We explicitly construct channels that achieve this information theoretical bound,
and establish that their optimality is metric-invariant, in that they achieve minimum leakage
with respect to \emph{any} choice of entropy that satisfies three mild conditions:
\emph{core-concavity} (which we define), \emph{symmetry} and \emph{expansibility}.
Next, in Section \ref{sec:Gains}, we extend our framework to non-symmetric (gain-based) entropies, introduce a
generalization of $g$-leakage, and establish a natural extension of our
main result to this class of entropies (Proposition~\ref{Thm:gain-extension}) for diagonal
gain matrices.
Finally, in Section~\ref{sec:Numerics}, we numerically investigate the effect of the
maximum allowable size of the pre-images of observables, the choice of the entropy, comparison with the baseline of
uniform randomization, and the effect of the adversary's knowledge of the true prior distribution.

Our proofs follow non-trivial techniques that we believe will add to the theoretical toolbox of
the research community.
Despite the theoretical nature of this work, we envisage possible applications of our results in
fields such as side
channels countermeasures in the style of bucketing \cite{kopf2010vulnerability,kopf2009provably},
in privacy preserving mechanisms like crowd-based anonymity protocols \cite{reiter1998crowds},
(Geo)-location privacy \cite{khoshgozaran2009private,ardagna2007location,sankar2013utility}, or obfuscation-based
web searching \cite{gervais2014quantifying}, \textit{etc}.
Detailed investigation of these connections and potential practical implementations will be part of our future work.

\paragraph*{\textbf{Literature Review}}
Generalisation of entropies is also discussed in a large body of literature. These entropies include the R\'enyi family that generalises Shannon and Min-entropy  e.g. \cite{jizba2004world,renner2005simple,fehr2014conditional,iwamoto2013information,golshani2009some,teixeira2012conditional}, and the Sharma-Mittal \cite{Sharma1975New} family, that generalises R\'enyi and Tsallis entropies. However, some of the  entropies  with very clear operational interpretation like Guesswork falls outside of their scope. Our generalised entropies includes all of them as special cases.

A main line of research to distinguish from is the classical  context of secure communication and secrecy systems \cite{schieler2014rate,hayashi2017equivocations,iwamoto2013revisiting,beigi2014quantum,bai2015improved} (e.g., in a wiretap setting), secure key distribution \cite{biondi2015attainable}, or steganography \cite{petitcolas1999information,moulin2003information}, \textit{etc.}, where the main goal is to reliably communicate secrets to a recipient while leaking the least to a third party. In contrast, in our setting, there is no intention to communicate any information at all as there is no intended recipient. There is instead a system that seeks to emit the least information to any outsider. That  said, similar to this line of work, the results in our paper is also information-theoretic, in that, our guarantees do not rely on computational difficulty of  certain operations (e.g. discrete logarithm) as in non-information-theoretic cryptography.

The general setting of information leakage outside of the communication setting has been studied in the quantitative information flow (QIF) literature \cite{heusser2010quantifying,doychev2015cacheaudit,mciver2015abstract,alvim2014additive}, works on private information retrieval (PIR) \cite{chor1998private} and private search queries \cite{domingo2009h,gervais2014quantifying}, as well as research on privacy-utility trade-offs \cite{khoshgozaran2009private,ardagna2007location,sankar2013utility}.

Particularly important from the field of QIF are advances on fundamental security
guarantees of leakage measures   (what security can be achieved)
and robust techniques  and results (how much a technique or
result is valid across different notions of leakage). However, most of the theoretical effort has been focused on analysing a given system as opposed to a design problem.

In the context of PIR, \cite{chor1998private} showed that in the presence of a single database,  the only information-theoretically private method is the trivial but unacceptable solution of requesting the entire database for each query. If multiple non-communicating replicas of the same database exist, then information-theoretic perfect privacy is achievable at a communication overhead. 
A heuristic work in the context of privacy in using internet search engines is \cite{domingo2009h}, where it is proposed that each search query should be accompanied by a number of bogus queries with similar frequencies as a means of camouflaging the real one. However, no analysis or claim about the optimality is produced.

The works on privacy-utility trade-off, e.g., in the context of location privacy \cite{shokri2015privacy,ardagna2007location,khoshgozaran2009private}, share a conceptual theme with our paper.
In particular, the cap on the size of the pre-image can be seen as a stylized constraint to achieve a minimum utility. However, in contrast to these works that only provide a methodology of finding a solution or an approximate solution, e.g., solving convex optimizations, we explicitly derive both the minimum leakage (exact, not a bound) and the optimal solution that achieves it. Moreover, there is no other work that considers our notion of robustness, i.e., measure-invariance.

Another line of research that is in the general spirit of utility-privacy trade-off is of $\epsilon$-differential privacy \cite{dwork2014algorithmic ,alvim2011differential}. There are a number of differences between  our approach and differential privacy, in terms of context, differential privacy is mostly for non-identifiability of individuals in published statistical data; in terms of implementation approach, the differential privacy is typically achieved by adding a controlled noise to the data, while in our setting, we conflate a controlled number of inputs by mapping them to the same output. But most significantly, there is a fundamental difference between information theoretic metrics of leakage and differential privacy: while information theoretic metrics rely on statistical averages, the differential privacy is a much stronger per realization metric.
Indeed, for instance as is shown by \cite{alvim2011differential}, differential privacy implies a bound on the min-entropy leakage but not vice-versa. 

There are some notable works that consider the problem of privacy-utility trade-off (PUT) from an information-theoretic point of view \cite{sankar2013utility,issa2016operational,liao2018privacy}. 
The focus of Maximal leakage metrics family is to provide a measure that is robust against the relative ``advantage'' each secret carries for the adversary, where this advantage may be unknown, i.e., robustness against the interest of the adversary in the secret. In contrast, our information-measure-invariance is about another robustness measure, that of modelling the attack mechanism: e.g. min-entropy is modelling an adversary that gets to make only one (best) guess after his observation; Guesswork, an adversary that can make sequential best guesses; Shannon, an adversary that can ask set-membership questions, etc.
This notion of robustness is arguably more relevant in a security context, where the ``value'' of the secret is known, but the capabilities of the attacker is not. 
In that sense, the original Maximal leakage \cite{issa2016operational} falls under an adversary that makes a single best guess (albeit about a function of the secret, as opposed to the secret directly). Although the generalization of that to $\alpha$-Maximal leakage and $f$-divergence allow other entropies, but still in the definition, the robustness is against the interest (or the advantage) of each secret to an adversary, and the choice of $\alpha$ or $f$ is fixed. Some of the relations of these notions are explored in \cite{issa2018operational}. Moreover, unlike ``capacity'' measures that consider worst case secret distribution for leakage, the secret distribution is a given in our setting. Finally, the focus of most the literature is on providing a robust measure to analyse a given channel,  as opposed to the constrained ``design'' problem, which is the focus of this work.

\section{Model}\label{sec:Model}

We will denote sets, random variables and realizations with calligraphic, capital, and small letters respectively, e.g. $\mathcal{X}$, $X$, $x$. We will denote the cardinality of a set $\XX$ by $|\XX|$. For a vector $p$, we use $p_{[i]}$ to denote the $i$'th largest element of $p$ where ties are broken arbitrarily. Also, we will use the notation $\|p\|_\alpha$ for the $\alpha$-norm of vector $p$, that is,  
$\|p\|_\alpha:=\left(\sum_{i=1}^np_i^\alpha\right)^{1/\alpha}$. The limit case of $\infty$-norm is $\|p\|_{\infty}:= p_{[1]}$.

Let $X$ represent the \emph{secret} as a discrete random variable that  can take one of the $n$ possibilities from $\XX:=\{1,\ldots,n\}$ with the (categorical) distribution of $p_{X}$. 
We assume that $p_X$ is publicly known and hence, we will refer to it as  \emph{the prior}.
For the rest of the paper, as is the convention, we will omit the superscript $X$ whenever not ambiguous and simply use $p(x)$ to refer to $p_X(x)$. Also, with a slight abuse of notation, we may use $p$ to refer to the vector of probabilities as opposed to its function form, that is, $p=(p(1),p(2),\ldots, p(n))$. The distinction should be clear from the context, e.g., when $p$ is used as the argument of a function with a vector input. 
Without loss of generality, assume that every secret has a strictly positive probability of realization, and that $p(x)$'s are sorted in non-increasing order, that is, $p(1)\geq p(2)\geq \ldots\geq p(n)>0$.

The system generates observables that can probabilistically depend on the secret. 
Let $\YY$ represent the discrete set of possible observables. Then the system can be modelled as a probabilistic discrete channel (henceforth referred to simply as channel) denoted by the triplet $(\XX,p_{Y|X},\YY)$, where $\XX$ and $\YY$ are the \emph{input} and \emph{output} \emph{alphabets} respectively, and $p_{Y|X}(y|x)$ for $x\in\XX$, $y\in\YY$ denotes the conditional probability distribution, i.e., the transition matrix. Specifically, it needs to satisfy the following (omitting the subscript for brevity, henceforth):
\begin{subequations}\label{eq:channel_requirements_distribution}
\begin{align}
&p(y|x)\geq0 & &\forall x\in\XX,\ 
y\in \YY;\label{subeq:channel_requirement_positive} \\
&\textstyle\sum_{y\in \YY}p(y|x)=1&  
& \forall x\in\XX.\label{subeq:channel_requirement_sum_to_1}
\end{align}
\end{subequations}
In the rest of the paper, we use the term channel to refer only to the conditional probability distribution (transition matrix), and will use the terms secret and input, as well as  observables and outputs interchangeably.
\hide{The distribution over the outputs can be computed by marginalization:
\begin{align*}
p(y) = \sum_{x\in\XX} p_x p(y|x), & & & y\in\YY
\end{align*} }
For a $y_0\in\YY$, we can define its pre-image as the subset of inputs that could have produced $y_0$ with non-zero probability. Formally, $\preimage(y_0):=\{x\in\XX:p(y_0|x)>0\}$.

The general setting in our paper is the following: an adversary observes the output of the channel and wants to infer about its input. The defender has a limited flexibility in designing the channel (the transition probability) and cannot change the prior, and wants to minimize the amount of information that the adversary can infer about the input by observing the outputs, i.e., leakage of information.

In the absence of any channel design constraint, one trivial solution that guarantees zero leakage is the following: showing the same output, say $y_0\in\YY$ for any realization of the input. More generally, any channel matrix that has the same rows will also lead to zero leakage. However,  such solutions might not be practical nor desirable in many areas of interest.
Technically, the property that enables the above trivial solutions is that the pre-image is allowed to be the entire space of the input, as $\preimage(y_0)=\XX$, and in particular,  $|\preimage(y_0)|=|\XX|=n$, where $y_0$ is any output whose entry in the identical rows is non-zero. Indeed, the problem of designing minimal-leakage channel becomes immediately non-trivial if we impose a cap on the size of the pre-images. That is, if we require that,   $\forall y_0\in \YY$, $|\preimage(y_0)|\leq k$ where $ k< n$. This constitutes the main setting of this paper.

To find leakage-minimal channels, we need a metric/measure to evaluate/quantify the information leakage. At a high level, this can be quantified as the difference between the prior uncertainty of an adversary and its posterior uncertainty, i.e., the uncertainty of the adversary about the input after observing the output of the channel -- on average. 
The prior uncertainty about the random variable $X$ is measured through its \emph{entropy}, and is denoted by $H(X)$ or simply $H(p)$. 
The posterior uncertainty is measured by \emph{posterior entropy} or the \emph{conditional entropy} of the random variable $X$ given the random variable $Y$. This is sometimes also referred to as the \emph{equivocation},  and is denoted by $H(X|Y)$. The classical choice for entropy and posterior entropy are the \textit{(Gibbs)-Shannon}'s:
\begin{subequations}\label{eq:Shannon_Entropy}
\begin{gather}
H(X) = -\sum_{x\in\XX} p(x)\log(p(x))\\
H(X|Y)=-\sum_{y\in\YY^+} p(y) \sum_{x\in\XX} p(x|y)\log(p(x|y))
\end{gather} 
\end{subequations}
where  $\YY^+$ is the set of outputs that have a strictly positive probability of realization, that is, $\YY^+=\{y\in\YY\mid \exists x\in\XX, p(y|x)>0\}$.
Also, $p(y)$ is the (total) probability that 
$y$ is observed by the adversary, and $p(x|y)$ is the 
posterior probability of the secret $x$ given that $y$ is observed, as given by the \emph{Bayes' rule}.
Specifically, $p(x|y)={p(x,y)}/{p(y)}={p(x)p(y|x)}/p(y)$ 
where $p(y)=\sum_{x'\in\XX}p({x'})p(y|x')$.

Shannon's entropy is related to the shortest coding of a random variable, which is also related to the average number of set-membership questions of an optimal adversary before getting to the value of a random variable.  However, this may not be a suitable measure in many contexts of interest \cite{massey1994guessing,smith2009foundations}, e.g., when the adversary needs to make a best guess in one, or multiple tries. For these operational scenarios, other more relevant entropies are introduced. For instance, the \textit{1-guess-error-probability}, defined as $H(X)=1-\|p\|_{\infty}=1-p_{[1]}$, is the probability that the best guess of an adversary about the secret is incorrect.   A closely related measure is the  \textit{Min-entropy}: $H(X)=-\log \|p     \|_{\infty}$. 
The \textit{$l$-guess-error-probability} extends to the cases where an adversary can submit $l$ best guesses, then    $H(X)=1-\sum_{i=1}^lp_{[i]}$ is the probability that none of them would be correct.  
Another frequently used entropy with a clear operational interpretation is \textit{guesswork (guessing)} entropy:  $H(X)=\sum_{i=1}^nip_{[i]}$. This measures the expected number of steps that takes a sequentially guessing optimal adversary to get to the secret. 
Another example is \textit{R\'enyi}, which is in fact a family of entropies parametrised by $\alpha\geq 0$, $\alpha\neq 1$, defined as:
\begin{align*}
H_{\alpha}(X)&=\frac{1}{1-\alpha}\log(\sum_{x\in\XX}(p(x))^{\alpha})\text{, or equivalently, } \\
H_{\alpha}(X)&=\frac{\alpha}{1-\alpha}\log\|p\|_{\alpha}
\end{align*} 
R\'enyi entropies can recover Shannon and Min-entropy  as limit cases 
by respectively letting $\alpha\to 1$ and $\alpha\to\infty$. For  $\alpha=2$, i.e., $H_2(X)=-\log\sum_{x\in\XX}(p(x))^2$, it is specifically called the \textit{collision} entropy. Likewise, the case of $\alpha=0$, i.e., $H_0(X)=\log|\supp(p)|=\log(n)$ is also known as the  \textit{Hartley} entropy.

For each of the aforementioned entropies, a posterior entropy can be defined in a meaningful way. For instance, the posterior $l$-guess-error-entropy ($1\leq l\leq n$) can be simply defined as the average failure rate of an adversary that makes a best guess about the secret after seeing the observable:
\begin{gather}\label{eq:l-Guess_Entropy_Conditional} H(X|Y)=\sum_{y\in\YY^+}p(y)\big(1-\sum_{i=1}^{l}\left(p_{X|y}\right)_{[i]}\big)
\end{gather}
 where $p_{X|y}$ is the vector of posterior probabilities given $Y=y$, i.e.,  
$p_{X|y}:=(p(x|y))_{x\in\XX}$. We are using the vector interpretation of probability distributions as it greatly simplifies the exposition.
Similarly, with respect to guesswork, we can write:
\begin{gather}\label{eq:Guesswork_conditonal} 
H(X|Y)=\sum_{y\in\YY^+}p(y)\big(\sum_{i=1}^ni(p_{X|y})_{[i]}\big).
\end{gather}

For the R\'enyi family, there is no universally accepted definition of its conditional form (e.g. \cite{jizba2004world,renner2005simple,fehr2014conditional,iwamoto2013information,golshani2009some,teixeira2012conditional}). Some of the candidates for the posterior R\'enyi entropy in the literature are:
\begin{subequations}
    \begin{align}
        H_{\alpha}(X|Y) =& \sum_{y\in\YY^+} p(y) H_{\alpha}(p_{X|y})\label{subeq:Renyi_cond_def_ng_1}\\
    H_{\alpha}(X|Y) =& H_{\alpha}(XY)-H_{\alpha}(Y) \notag\\ 
    & =\frac{1}{1-\alpha} \log \left(\frac{\sum_{x,y}\left(p(x,y)\right)^{\alpha}}{\sum_{y}\left(p(y)\right)^{\alpha}}\right)\label{subeq:Renyi_cond_def_ng_2}\\
    H_{\alpha}(X|Y) =& \frac{1}{1-\alpha}\max_{y\in\YY^+} \big(\log 
 \|p_{X|y}\|^{\alpha}_{\alpha}\big)\label{subeq:Renyi_cond_def_ng_3}\\
    H_{\alpha}(X|Y) =& 
\frac{\alpha}{1-\alpha}\log\big(\!\sum_{y\in\YY^+}\!p(y)\|p_{X|y}\|_{\alpha}\big) 
\label{subeq:Renyi_cond_def_alpha}  \\ 
    H_{\alpha}(X|Y) =& \frac{1}{1-\alpha} \log 
    \big(\!\sum_{y\in\YY^+}\! p(y) \|p_{X|y}\|^{\alpha}_{\alpha}\big)\label{subeq:Renyi_cond_def_alpha_alpha}\\
   H_{\alpha}(X|Y) =& 
   -\log\big(\!\sum_{y\in\YY^+}\!p(y)\|p_{X|y}\|^{\frac{\alpha}{\alpha-1}}_{\alpha}\big)
   \label{subeq:Renyi_cond_def_alpha_other}
   \end{align}
\end{subequations}
In particular, definition \eqref{subeq:Renyi_cond_def_ng_1} is introduced in \cite[eq. (2.15)]{cachin1997entropy}, definition \eqref{subeq:Renyi_cond_def_ng_2} in \cite[eq. (2.9)]{golshani2009some} and \cite[eq.(2.17)]{jizba2004world},   definition \eqref{subeq:Renyi_cond_def_ng_3} in \cite[Sec. 2.1]{renner2005simple} by setting $\epsilon=0$ in their conditional $\epsilon$-smooth R\'enyi
entropy definition. It is shown (e.g. in \cite[Theorem 7]{teixeira2012conditional}) that none of the definitions \eqref{subeq:Renyi_cond_def_ng_1}-\eqref{subeq:Renyi_cond_def_ng_3} satisfy the basic property of monotonicity, i.e., conditioning  reduces entropy (CRE) in general. That is, for each of these definitions, one can find joint distributions on $X,Y$ such that $H_{\alpha}(X|Y)>H_{\alpha}(X)$.
This makes them rather unsuitable for our setting: we make the assumption that the average uncertainty of the adversary about the input should not increase after observing the output of a channel, based on the argument that the adversary always has the option of simply ignoring his observation. Therefore, we only consider the definitions \eqref{subeq:Renyi_cond_def_alpha} to \eqref{subeq:Renyi_cond_def_alpha_other}, which, as we will show satisfy the data processing inequality (DPI), which in part implies CRE.
\eqref{subeq:Renyi_cond_def_alpha} is the recognised Arimoto definition of conditional R\'enyi \cite{Arimoto1975Information}. Definition~\eqref{subeq:Renyi_cond_def_alpha_alpha} is proposed in \cite[Sec. II.A]{hayashi2011exponential} and \eqref{subeq:Renyi_cond_def_alpha_other} is introduced in \cite{fehr2014conditional}. We also note that  \eqref{subeq:Renyi_cond_def_alpha_alpha}, \eqref{subeq:Renyi_cond_def_alpha}
are respectively equivalent to $H_{1+s}$ and $H^{\uparrow}_{1+s}$ defined in eq. (15) and (16) in \cite{hayashi2017equivocations} by taking $\alpha=1+s$. 

\subsection{Introducing a generalised entropy}
In this paper, we consider a generalized entropy that encompass all of the above cases. In particular, it has the following structure:
\begin{equation}
\label{eq:cond_entr_gen}
H(X\mid Y)=\eta\Big(\!\sum_{y\in\YY^+}p(y)
F\left(p_{X \mid y}\right)\Big),
\end{equation}
where $\eta$ is just an $\RR\to\RR$ function, and $F$ is a bounded scalar function over the space of probability distributions with the following properties:
\begin{compactitem}
	\item\emph{symmetry}, i.e., its value only depends on the shape of a distribution and does not change with any re-ordering of the probabilities (re-labelling the random variables);
	\item\emph{expansibility}, i.e., its value does not change by padding the probability distribution with zero entries;
\end{compactitem} 
Moreover, one of the following two conditions holds (a property that we just call \emph{core-concavity}):\footnote{Note that only case (a) can be considered as the definition, as case (b) can be transformed to case (a) by $F'(p)=-F(p)$, $\eta'(x)=\eta(-x)$.}
\begin{subequations}\label{eq:specs_of_g_and_F}
	\begin{align}
	&\text{$\eta$: increasing, and $F$: concave; or}\label{eq:subeq:g_inc_F_concave}\\  
	&\text{$\eta$: decreasing, and $F$: convex.}\label{eq:subeq:g_dec_F_convex}
	\end{align}
\end{subequations}  
By definition, $F(p)$, as a scalar function with vector arguments, is concave (respectively, convex) in $p$ iff:  $\forall \lambda \in[0,1]$ and for any
    probability distributions $p_1, p_2$ over $\XX$, we have: $\lambda F(p_1)+(1-\lambda) F(p_2)\leq$ 
    (respectively, $\geq$) $F(\lambda p_1+(1-\lambda)p_2)$. 

Note that the form of the conditional entropy in \eqref{eq:cond_entr_gen} governs the form of the unconditional entropy as well (e.g. by taking $Y$ and $X$ to be independent). Specifically, \begin{gather*}
H(X)=H(p)=\eta\left( 
F\left(p\right)\right).
\end{gather*}

For cases where $\eta(\cdot)$ is strictly monotonic, our generalised conditional entropy can be re-written in terms of the non-conditional entropy as follows:
\begin{gather*}
H(X|Y) = \eta\left(\sum_{y}p(y)\eta^{-1}\left(H(p_{X|y})\right)\right),
\end{gather*}
This gives another interpretation for $H(X|Y)$ as the Kolmogorov-Nagumo  average of the unconditional entropy with respect to function $\eta^{-1}(\cdot)$ (see e.g. \cite{furuichi2012mathematical}).

\begin{prop}\label{prop:encompasses_entropies}
All of the conditional entropies: $l$-guess-error probability, Guesswork \eqref{eq:Guesswork_conditonal} and R\'enyi entropies according to \eqref{subeq:Renyi_cond_def_alpha}--\eqref{subeq:Renyi_cond_def_alpha_other} (which includes Shannon \eqref{eq:Shannon_Entropy} and Min-Entropy as limit cases)  are special cases of our generalised definition in \eqref{eq:cond_entr_gen}.
\end{prop}

\begin{IEEEproof} 
Shannon entropy can be represented by taking $\eta$ to be the identity function, i.e., $\eta(x)=x$, and $F(p)=-\sum_{i=1}^np_i\log(p_i)$ which is well known to be a symmetric concave function over the space of probability distributions, and also expansible with the convention of $0\log0=0$. Likewise, for $l$-guess-error probability and guesswork, $\eta$ can be taken as the identity function as well. The $F(p)$ will be $1-\sum_{i=1}^lp_{[i]}$ and $\sum_{i=1}^nip_{[i]}$, respectively, which are again known to be concave.  For the Arimoto conditional R\'enyi entropy as in~\eqref{subeq:Renyi_cond_def_alpha}, 
we can take  $\eta(x)=\frac{\alpha}{1-\alpha} \log (x)$ on $\RR^+$ and $F(p)=\|p\|_{\alpha}$. For the conditional R\'enyi entropy as per \eqref{subeq:Renyi_cond_def_alpha_alpha}, we can take 
$\eta(x)=\frac{1}{1-\alpha} \log (x)$ on $\RR^+$ and $F(p)=\|p\|^{\alpha}_{\alpha}
=\sum_{i=1}^np_i^\alpha$. For both cases, $F$ is a symmetric function. Moreover, when $0\leq \alpha<1$, $\eta$ is increasing and $F$ is concave, 
and when $\alpha>1$, $\eta$ is decreasing and $F$ is convex.  
For definition~\eqref{subeq:Renyi_cond_def_alpha_other}, we can take $\eta(x)=-\log(x)$, which is a decreasing function, and $F(p)=\|p\|^{\frac{\alpha}{\alpha-1}}_{\alpha}$, which is a convex function for any $\alpha\geq 0$.
\end{IEEEproof}

\paragraph*{Remark} Another important  family of entropies is the Sharma-Mittal  parametrised entropies \cite{Sharma1975New} defined as:
\begin{gather}\label{eq:Sharma-Mittal}
H_{\alpha,\beta}(X) = \frac{1}{\beta-1}\left(1-\left(\|p\|^{\alpha}_{\alpha}\right)^{\frac{1-\beta}{1-\alpha}}\right), \quad \alpha\geq0,\ \alpha,\beta\neq 1.
\end{gather}
This family can retrieve R\'enyi as $H_{\alpha,\beta\to 1}(X)$ (including Shannon as $H_{\alpha\to 1,\beta\to 1}(X)$), as well as Tsallis entropies \cite{tsallis1988possible}: $H_{\alpha,\alpha}(X)=\frac{1}{1-\alpha}\left(1-\|p\|^{\alpha}_{\alpha}\right)$. 
$H_{\alpha,\beta}(X)$ also is particular case of our generalised entropies. This can be seen, for instance, by taking $\eta(x)=\frac{1}{\beta-1}(1-x^{\frac{1-\beta}{1-\alpha}})$ and $F(p)=\|p\|^{\alpha}_{\alpha}$.
For $\alpha>1$ and any $\beta\neq 1$, $\eta(x)$ is decreasing and $F(p)$ is convex, and for $0<\alpha<1$ and any $\beta\neq 1$, $\eta(x)$ is increasing and $F(p)$ is concave. As with the R\'enyi entropy, there is no generally agreed-upon conditional form of the Sharma-Mittal entropies.  Our generalised form allows multiple candidates. If we take the same $\eta$ and $F$ functions as above, we get the following form of conditional Sharma-Mittal entropy:
\begin{gather*}
H_{\alpha,\beta}(X|Y)=\frac{1}{\beta-1}\left(1-\left(\sum_{y\in\YY^+}p(y)\|p_{X|y}\|^{\alpha}_{\alpha}\right)^{\frac{1-\beta}{1-\alpha}}\right)
\end{gather*}
With the above definition, the limit $H_{\alpha,\beta\to 1}(X|Y)$ retrieves the Arimoto's form of conditional entropy as in \eqref{subeq:Renyi_cond_def_alpha}. Different choices of $\eta$ and $F$ are possible which result in alternative forms of the conditional entropies.

\paragraph{Derived generalised Information Theoretical Measures}
Given our generalised entropy and conditional entropy $H(X), H(X|Y)$ we can define a generalization of
mutual information: this is defined as the difference between the prior (unconditional) and posterior (conditional) entropies:
\begin{gather*}
I(X;Y) = H(X)- H(X|Y)
\end{gather*}
Arimoto's $\alpha-$mutual information \cite{Arimoto1975Information} is a particular case of the above.
In our setting mutual information is synonymous with leakage: it quantifies, according to a chosen entropy $H$ the reduction in uncertainty of an attacker given the observations.
Building on the generalized mutual information we can also generalize the
channel capacity: this is the maximum mutual information where the maximization is with respect to all distributions over $X$:
\begin{gather*}
C(X;Y) = \max_{p_X} I(X;Y) 
\end{gather*}
Min-capacity \cite{smith2009foundations} and Maximal leakage \cite{issa2016operational,liao2017hypothesis} are particular examples of the channel capacity by choosing the underlying entropy to be  Min-Entropy.

\paragraph{ Symmetry, core-concavity, \emph{majorization} and \emph{Schur-concavity}}

The symmetry and core-concavity properties together have an intuitive implication:
that the distributions that are ``closer to uniform'' represent a higher entropy. 
This is formalized through the notions of 
\emph{majorization} and \emph{Schur-concavity}, which we will use in our proofs. Here, we provide a brief overview:  
For vectors $a$, 
$b\in \mathbb{R}^n$, we denote $a\succ b$ and say 
$a$ \emph{majorizes}  $b$ (or $b$ is \emph{majorized} 
or \emph{dominated} by $a$)  iff:
$\sum_{i=1}^j a_{[i]}\geq \sum_{i=1}^jb_{[i]}$ for all $j=1,\ldots,(n-1)$, 
and $\sum_{i=1}^na_i=\sum_{i=1}^nb_i$. 
For probability distributions, $p_1\succ p_2$ implies that $p_1$ is further away (more skewed away) from uniform distribution compared with $p_2$.

A function $f:\mathbb{R}^n\to \mathbb{R}$ is called \emph{Schur-concave} iff: 
for $a, b\in \mathbb{R}^n$, $a\succ b$ implies 
$f(a)\leq f(b)$. In words, the value of a Schur-concave function (over the space of probabilities) increases as its input gets closer to the uniform distribution.
A \emph{Schur-convex} function is defined 
likewise where the last inequality is flipped.
A basic result in convex analysis (see e.g. 
\cite[Prop. 3.C.2]{marshall2010inequalities}) states that: 
Any function that is symmetric 
and concave (convex, resp.) is also Schur-concave (Schur-convex, resp.).  
Therefore, symmetry and core-concavity conditions imply that our entropy functions  
are Schur-concave as well.

As mentioned before the information leakage can be quantified as the mutual information $I(X;Y)=H(X)-H(X|Y)$, a quantity which we want to minimize.  

As we already argued, Shannon entropy may not be a suitable measure for many contexts of interest, which motivated introduction of other entropies and leakage measures.
A main concern is which one to choose for the problem of leakage-minimal design, especially as each entropy has a distinct operational interpretation, and most awkwardly, some depend on modelling the behaviour/abilities of the adversaries. A desirable property would be to have a solution that is invariant under the choice of the entropy, i.e., a channel that would simultaneously minimize the leakage for any reasonable choice of the entropy, if such a solution exists. This is exactly the goal of this paper. Hence, we express the problem statement of our paper as follows:

\begin{framed}
    \begin{align*}
    \textbf{Given:}& & & \parbox{19em}{$p_X=(p_1,\ldots,p_n)$ in non-increasing order, and $k<n$} \\
    \textbf{Goal:}& & & \parbox{19em}{Find $p_{Y|X}$ that minimizes $H(X)-H(X|Y)$ for any choice of entropy}
    \\
    & & & \text{subject to:}\qquad |\preimage(y)|\leq k\ \forall y\in\YY.
    \end{align*}
\end{framed}
Note that in our setting, $H(X)$ is fixed, and hence, the above minimization can be equivalently expressed as maximization of $H(X|Y)$.

\subsection{Basic properties of our generalized entropies and leakage}
Here, we show that our generalized entropies in \eqref{eq:cond_entr_gen} satisfy some desirable properties, namely, non-negativity of the leakage and the \emph{data-processing inequality}.
\begin{prop}\label{prop:leakage_positive_dpi}
    Any generalized entropy as defined in \eqref{eq:cond_entr_gen} satisfies:
    \begin{compactenum}[(a)]
        \item Non-negativity of leakage, defined as $H(X)-H(X|Y)$; and
        \item Data processing inequality (DPI): consider random variables $X$, $Y$, $Z$, and assume that given $Y$, 
        $Z$ is conditionally independent from $X$ (sometimes denoted as $X\rightarrow Y\rightarrow Z$). Then for any entropy measure in \eqref{eq:cond_entr_gen}, we have: $H(X|Z)\geq H(X|Y)$. 
    \end{compactenum} 
\end{prop}    
\begin{IEEEproof}
    Part~(a) follows as a special case of part~(b) if we take $Z$ to be independent from $X$.  Hence, we just prove part~(b): Referring to \eqref{eq:cond_entr_gen}, 
    we have: \begin{multline*}H(X|Z)=\eta\left(\sum_{z}p(z)
    F\left(p_{X|z}\right)\right)=\\
    \eta\left(\sum_{z}p(z)F
    \left(\sum_{y}p(y|z)p_{X|y,z}\right)\right),
    \end{multline*} 
    where we used $p_{X|z}=\sum_{y}p(y|z)p_{X|y,z}$. 
    Next, note that for any given  $z$, $p(y|z)$ constitute convex coefficients, since they are non-negative for each $y$, and $\sum_{y}p(y|z)=1$. 
    Therefore, following Jensen's inequality, for both cases \eqref{eq:subeq:g_inc_F_concave} and~\eqref{eq:subeq:g_dec_F_convex}, 
    we have: 
   \begin{gather*}
   H(X|Z)
   \geq \eta\left(\sum_{z}\sum_{y}p(z)p(y|z)F
   \left(p_{X|y,z}\right)\right).
   \end{gather*}
   The conditional independence of $Z$ and $X$ given $Y$ means:   $p_{X|y,z}=p_{X|y}$.  Hence: 
   \begin{multline*}
   H(X|Z)
   \geq \eta\left(\sum_{y,z}p(y,z)F
   \left(p_{X|y}\right)\right)=\\
   \eta\left(\sum_{y}p(y)F
   \left(p_{X|y}\right)\right)=H(X|Y).
   \end{multline*}
\end{IEEEproof}

Our data processing inequality (DPI) applies to generalized conditional entropies in \eqref{eq:cond_entr_gen}. In particular, it recovers similar results  in \cite{fehr2014conditional,iwamoto2013revisiting} for conditional R\'enyi in the forms of \eqref{subeq:Renyi_cond_def_alpha}, \eqref{subeq:Renyi_cond_def_alpha_alpha} as special cases.

\hide{Specifically, 

\begin{prop}\label{Prop:preimage}
    Consider the problem of designing a channel given a distribution $\PP$ over the 
    set of inputs (secrets) $\SecSet$ with the constraint that the pre-image of each observable 
    has to be limited to at most $k$ elements. Then there is an optimal channel construction with 
    respect to leakage in which no two observable have the same pre-image.   
\end{prop}
\begin{IEEEproof}
    Consider any optimal channel $C$ in which at least two observables, say $y_1$ and $y_2$ 
    share the same pre-image $M\subset \Theta$. Now, consider an alternative channel constructed 
    by merging $y_2$ into $y_1$, i.e., for all $\theta\in M$,
    $C_{\mathrm{new}}[\theta,y_1]=C[\theta,y_1]+C[\theta,y_2]$ and    
    $C_{\mathrm{new}}[\theta,y_2]=0$. Clearly, $C_{\mathrm{new}}$ represents a valid channel 
    (entries are still non-negative and rows add up to one).   
    Moreover, $C_{\mathrm{new}}$ leaks less information than $C$. 
    The latter follows from a \red{straightforward} application of the following lemma 
    (a generalization of the data processing inequality for our generic entropies):
}

\section{Analysis}\label{sec:Analysis}
The first point to observe is that in our setting, the prior $p_X$ is a given parameter and the choice of the channel does not impact the prior uncertainty $H(X)$. Hence, the objective of minimizing the leakage becomes equivalent to maximizing the posterior entropy.
In this section, we derive (in closed form) the maximum possible posterior entropy that can be 
achieved among all feasible channels for a given prior $p$, 
a pre-image size cap $k$, and a measure of entropy $H$  (Theorem~\ref{Thm:main}-A). Our result is constructive, 
in that, in Algorithm~\ref{Alg:Universal_Optimal_Cloaking}, we explicitly provide a channel that 
achieves this maximum posterior entropy (and hence, minimum leakage) for any symmetric, expansible, 
core-concave measure of entropy (Theorem~\ref{Thm:main}-B).
As we mentioned before, since each entropy measure has its own distinct form and interpretation, 
    it could have been the case that optimality of any channel sensitively depended on 
    the choice of entropy. The fact that such metric-invariant optimal channels exist  in our setting  is one of our contributions.

Before we present our formal result, let us develop a feeling about the behaviour of an optimal channel. 
Intuitively, the pre-images should be at the maximum allowed size of $k$, since the maximum number of inputs will be conflated with each other to increase the adversary's ambiguity. Also, 
intuitively, we should try to induce posterior distributions  over the pre-images 
that are as close to uniform distribution over $k$ elements as possible, since any well-defined 
measure of uncertainty increases as the  distributions gets closer to uniform. The ideal case is 
that given any shown output, after the Bayesian update, the input be equally likely any of the 
$k$ members of its pre-image.  However, if the prior distribution is too skewed and the cap size of 
the pre-images is small, then inducing uniform posteriors might not be feasible, as the inputs with too big prior probabilities 
will still have higher posteriors. If a prior probability of an input is too big to be made uniform 
in the posterior, i.e., a ``giant'', then it should be instead maximally leveraged against to hide  
other inputs in its ``shadow''. So, intuitively, an optimal channel should try  to 
induce posteriors that are uniform over as many of the small probability inputs as 
possible and the giants should always be included in the pre-image to provide coverage for the small-probability inputs.

In order to formally present our results, we need to introduce some auxiliary parameters. 
Given $k$ and $p=(p(1),\ldots,p(n))$, sorted in non-increasing order, let index $j^*$ be:
\begin{gather}	\label{eq:J_definition}
j^*:=\min\left\{j: 1\leq j\leq k, p(j)\leq \dfrac{\sum_{i=j}^{n}p(i)}{k-j+1}\right\}.
\end{gather}
Note that for $j=k$, the condition $p(j)\leq \sum_{j=1}^{n}p(i)/(k-j+1)$ reduces to 
$p(k)\leq  \sum_{i=k}^{n}p(i)$, which is trivially satisfied. Therefore, $j^*$ is well-defined 
(i.e., can always be found), and  we  have $1\leq j^*\leq k$. Along the lines of 
the above intuitive discussion, the first $j^*-1$ inputs are the giants.
Next, for a prior distribution $p=(p(1),\ldots,p(n))$ sorted in non-increasing order, cap-size $k$, and the corresponding $j^*$ given by \eqref{eq:J_definition}, 
let $\pi=(\pi_1,\ldots,\pi_k)$ denote the probability distribution over $k$ elements defined as follows:\begin{gather}
\pi:=\Big(p(1),\dotsc,p(j^*-1),\dfrac{\sum_{i=j^*}^np(i)}{k-j^*+1},\dotsc,\dfrac{\sum_{i=j^*}^np(i)}{k-j^*+1}\Big),
\text{ i.e.:}\notag\\
\pi_l\!=\!p(l)\!:\!\ l\!\leq\! j^*\!-\!1,\ \,\, 
\pi_l\!=\!\frac{\sum_{i=j^*}^np(i)}{k-j^*+1}\!:\!\ j^*\!\leq\! l\!\leq\! k \label{eq:pi_j_definition} 
\end{gather}
In words, $\pi$ is a $k$-sized probability distribution (in vector format) that is constructed by 
keeping the top $j^*-1$ probabilities of the prior as is, and then wrapping or mashing
the remaining probabilities of the prior together and spreading them evenly over the 
remaining $k-(j^*-1)$ elements.
Note that if $j^*=1$, then $\pi$ is simply the uniform distribution over the entire $k$ elements.

Finally, let $\Env(S)$, where
$S\subseteq\XX$ and $|S|\leq k$,    denote the set of subsets of $\XX$ that 
include all the elements of $S$ and have size equal to $k$. Formally, 
$\Env(S):=\{M\subset \XX:S\subseteq M,\left\vert{M}\right\vert= k\}$. Note that $\Env(\emptyset)$ is just the set of all $k$-sized subsets of $\XX$.
This notation is used in our Algorithm as well as our proofs.
For a simple example, suppose $\XX=\{1,2,3,4\}$ and $k=3$, 
then $\Env(\{1\})=\{\{1,2,3\},\{1,2,4\},\{1,3,4\}\}$, and $\Env(\{1,2\})=\{\{1,2,3\},\{1,2,4\}\}$, 
and so on.
We are now ready to express our main result:

\begin{thm}\label{Thm:main}
Let $p=(p(1),\ldots,p(n))$ be the prior (sorted in non-increasing order), and let $k$ be 
the maximum allowed size of the pre-images. 
Suppose the posterior entropy $H$ has the generic format of \eqref{eq:cond_entr_gen}.
Let  
the probability distribution 
$\pi$ be as described in 

\eqref{eq:pi_j_definition}. 
Then:
\begin{itemize}
\item[A.] The maximum achievable posterior entropy among all  
channels is $H(\pi)$.
\item[B.] Algorithm~\ref{Alg:Universal_Optimal_Cloaking} explicitly provides a feasible 
 channel that achieves the above maximum posterior entropy for any  
choice of our entropy functions, and is hence metric-invariant.
\end{itemize}  
\end{thm}
\begin{algorithm}[h!]
\caption{\footnotesize{Optimal channel for a given $p$, $k$ 
(Theorem~\ref{Thm:main})}}\label{Alg:Universal_Optimal_Cloaking}
\begin{framed}
\renewcommand{\algorithmicrequire}{\textbf{Input:}}
\renewcommand{\algorithmicensure}{\textbf{Output:}}
\begin{algorithmic}[1]
\Require{$p=(p(1),\ldots,p(n))$ in non-increasing order, $k$}
\Ensure{$p_{Y|X}$}
\State 	\textbf{Find} $j^*\gets\min\Big\{1\leq j\leq k: p(j)\leq \dfrac{\sum_{i=j}^np(i)}{k-j+1}\Big\}$ 
\State \textbf{Solve} 
$\displaystyle\!\!\!\sum_{M\in\Env(\{1,\ldots,j^*-1,i\})}\!\!\! v_M\,\,\, =\,\,\, p(i),\quad\forall i=j^*,\ldots,n$
\item[] \quad\textit{s. t.:}\ \  $v_M \geq 0, \quad\forall M\in\Env(\{1,\ldots,j^*\!-\!1\})$
\item[]
\State $p(y_M|i)\gets v_M/p(i)$\hfill  $\forall i=j^*,\dotsc, n$
\item[] \hfill$\forall M\in\Env(\{1,\ldots,j^*\!-\!1,i\})$
\State $p(y_M|i)\gets v_M(k-j^*+1)/\sum_{j=j^*}^np(j)$ \item[]\hfill$\forall i=1,\dotsc, j^*\!-\!1$ 	
\item[] \hfill$\forall M\in\Env(\{1,\ldots,j^*\!-\!1\})$
\State $p(y|x)\gets0$ \hfill \textit{everywhere else}
\end{algorithmic}
\end{framed}
\end{algorithm}

In the algorithm, each distinct pre-image is associated with a unique output. Consequently, each output is indexed by its associated pre-image.
Note that the 
        optimal channel may not be unique, since the set of solutions to the linear feasibility system in Step 2 of Algorithm~\ref{Alg:Universal_Optimal_Cloaking} are in general convex 
        polyhedra. The theorem guarantees that all of such solutions are optimal and their optimality is metric-invariant.

At its core, Algorithm~\ref{Alg:Universal_Optimal_Cloaking} is doing something simple: it 
generates a channel such that given any output $y$  shown to the adversary, the posterior distribution over the inputs in its pre-image 
is exactly $\pi$. It 
does so by \emph{always} including the inputs $1,\ldots,j^*-1$ in the pre-images,
and carefully choosing the randomization of the transition matrix such that the posterior probability 
over the remaining $k-(j^*-1)$ items of a pre-image is uniform (guaranteed by the solution of the 
linear system of equations in Step 2), and the posterior distribution over the first $j^*-1$ 
elements of the pre-image is exactly the first $j^*-1$ entries of the prior (guaranteed by Steps 3 and 4).  

Before we present the proof, let us 
compute the optimal channel for a few toy examples to gain 
some intuition. 
Consider the case $\XX=\{1,2,3,4\}$ and $k=3$. 
We have the following possible size 3 pre-images:
\[M_1\!=\!\{\!1,2,3\!\},M_2\!=\!\{\!1,2,4\!\},
M_3\!=\!\{\!1,3,4\!\},M_4\!=\!\{\!2,3,4\!\}\]
\begin{figure*}[ht]
    \centering
    \begin{subfloat}[][]{\label{Fig:p1_barplot}
    \includegraphics[width=0.32\textwidth,height=120pt]{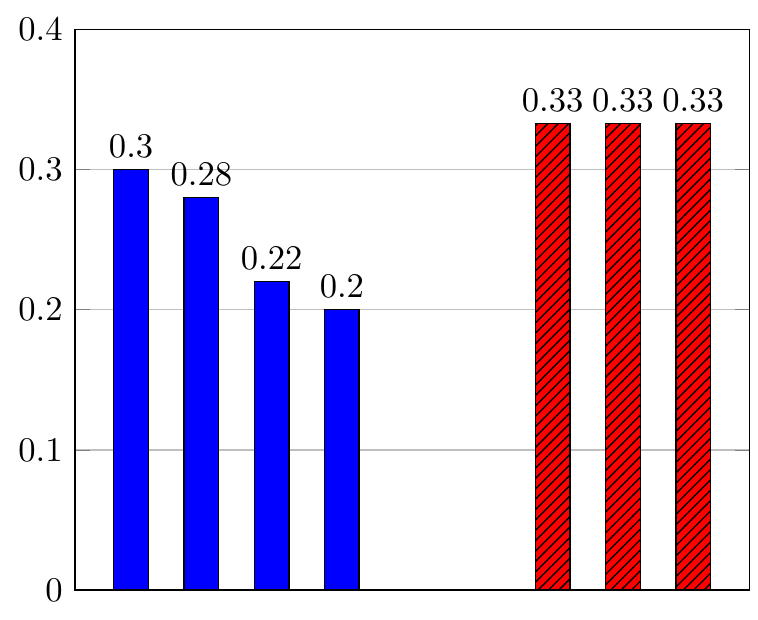}}
    \end{subfloat}\ 
    \begin{subfloat}[][]{\label{Fig:p2_barplot}
    \includegraphics[width=0.32\textwidth,height=120pt]{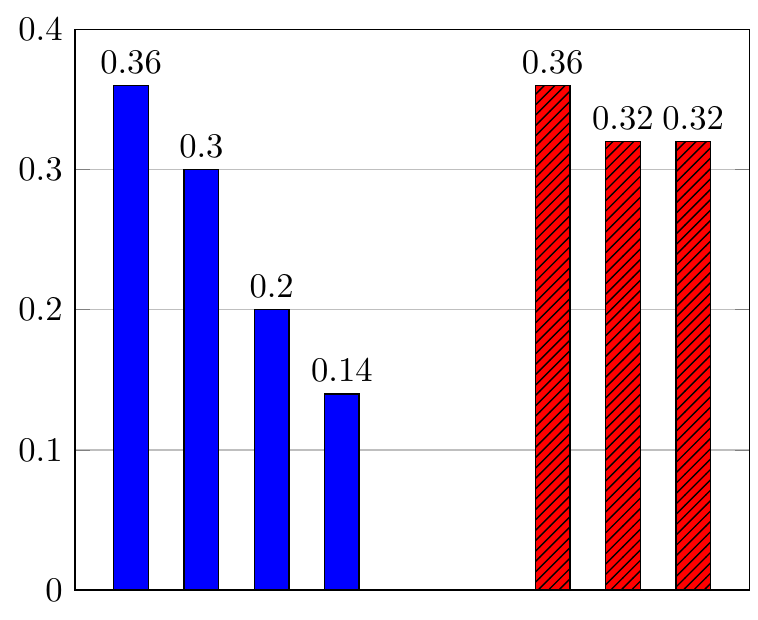}}
    \end{subfloat}\ 
    \begin{subfloat}[][]{\label{Fig:p3_barplot}
            \includegraphics[width=0.32\textwidth,height=120pt]{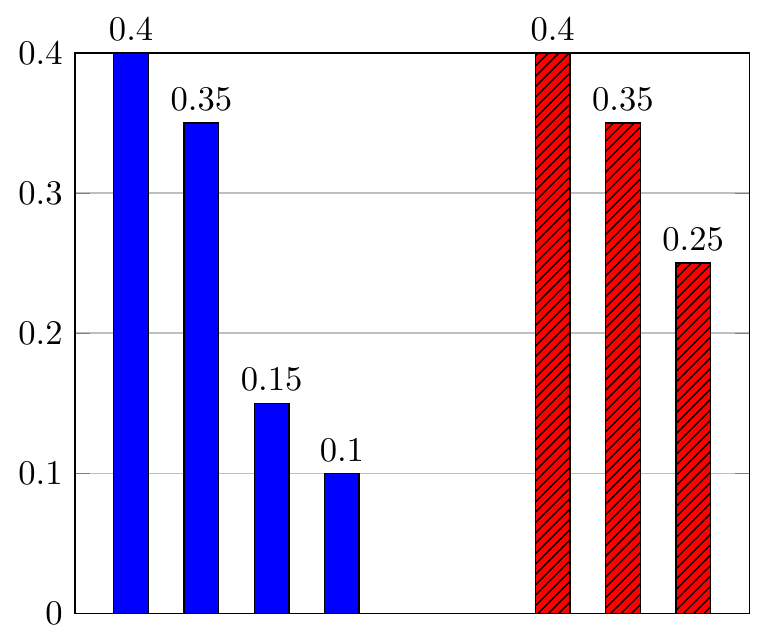}}
    \end{subfloat}
    \caption{Three toy examples for illustration of Theorem~\ref{Thm:main}. The priors $p_1$, $p_2$ and $p_3$ beside their corresponding $\pi$ (as described in the theorem) are respectively shown in (a), (b) and (c). Note that the priors are increasingly more skewed away from the uniform. In particular, in (a): we have $j^*=1$, i.e., no giants, in (b): $j^*=2$, i.e., one giant, and in (c): $j^*=3$, i.e., 2 giants.}
\end{figure*}

First, consider the following prior over 
the secrets:
$ p_1=(0.3,0.28,0.22,0.2)$.
We have: $p_1(1)=0.3\leq 1/k=1/3=0.33$, hence $j^*=1$,
and an optimal channel must induce  $\pi=(1/3,1/3,1/3)$ posterior distributions.
Since, $j^*=1$, the linear system in Step-2 of the algorithm 
is as follows:
\begin{gather*}
v_{\{1,2,3\}}+v_{\{1,2,4\}}+v_{\{1,3,4\}} = 0.3\\
v_{\{1,2,3\}}+v_{\{1,2,4\}}+v_{\{2,3,4\}} = 0.28\\
v_{\{1,2,3\}}+v_{\{1,3,4\}}+v_{\{2,3,4\}} = 0.22\\
v_{\{1,2,4\}}+v_{\{1,3,4\}}+v_{\{2,3,4\}} = 0.2\\
v_{\{1,2,3\}}, v_{\{1,2,4\}}, v_{\{1,3,4\}}, v_{\{2,3,4\}} \geq 0
\end{gather*}
which, after solving it and following Steps 3 and 4 of the algorithm 
yields the optimal channel as:
{\small
\begin{center}
\begin{tabular}{@{} C{40pt} C{30pt} C{30pt} C{30pt} C{30pt}@{}}%
\toprule%
& $y_{\{\!1,2,3\!\}}$\quad  & $y_{\{\!1,2,4\!\}}$\quad    &     $y_{\{\!1,3,4\!\}}$\quad      &    $y_{\{\!2,3,4\!\}}$\quad \\
\cmidrule(r){2-5}
(0.30) $1$:& 0.4444 & 0.3778   &   0.1778    &    0  \\
(0.28) $2$:&    0.4762   &  0.4048     &    0  &  0.1190\\
(0.22) $3$:&    0.6061     &     0  &  0.2424  & 0.1515  \\
(0.20) $4$:&       0   &  0.5667  &  0.2667 &   0.1667 \\
\bottomrule
\end{tabular}
\end{center}
}
We can check that the above optimal channel 
induces uniform posterior distribution over 3 elements of any pre-image.
Recall from Bayes' rule that   
$p(x|y)=p(x)p(y|x)/{p(y)}$. 
Since the denominator is the same for a given output, we just need to verify 
$p(x)p(y|x)$ 
is the same for all $x\in \preimage(y_M)=M$. For instance, for $M_1=\{1,2,3\}$ we have:
$0.3\times 0.4444= 0.28\times 0.4762=0.22\times 0.6061=0.1333$. Hence, $p(1|y_{1,2,3})=p(2|y_{1,2,3})=p(3|y_{1,2,3})=1/3$. Similarly, 
for $M_2=\{1,2,4\}$ we have: $
0.3\times 0.3778= 0.28\times 0.4048=0.2\times 0.5667=0.1133$. And finally, 
for $M_3=\{1,3,4\}$, we have: $
0.3\times 0.1778= 0.22\times 0.2424=0.2\times 0.2667=0.0533$. 

Now consider an alternative prior: $p_2=(0.36,0.3,0.2,0.14)$, 
We have: $p_2(1)=0.36>1/k$ but 
$p_2(2)=0.3\leq(0.3+0.2+0.14)/(k-1)=0.64/2=0.32$, 
therefore $j^*=2$, and the optimal channel will always 
include $1$ in the pre-images and induce
$\pi=\left(p_2(1),(p_2(2)+p_2(3)+p_2(4))/2,(p_2(2)+p_2(3)+p_2(4))/2\right)=(0.36,0.32,0.32)$ posterior distributions. The corresponding linear 
system in Step~2 is:
\begin{gather*}
v_{\{1,2,3\}}+v_{\{1,2,4\}} = 0.3\\
v_{\{1,2,3\}}+v_{\{1,3,4\}} = 0.2\\
v_{\{1,2,4\}}+v_{\{1,3,4\}} = 0.14\\
v_{\{1,2,3\}}, v_{\{1,2,4\}}, v_{\{1,3,4\}} \geq 0
\end{gather*}
which yields the optimal channel as:
{\small
\begin{center}
\begin{tabular}{@{} C{40pt} C{30pt} C{30pt} C{30pt} @{}}
\toprule
& $y_{\{\!1,2,3\!\}}$\quad  & $y_{\{\!1,2,4\!\}}$\quad    &     $y_{\{\!1,3,4\!\}}$\quad      \\   
\cmidrule(r){2-4}
(0.36) $1$:& 0.5625  &  0.3750   & 0.0625       \\
(0.30) $2$:& 0.6000      & 0.4000      &   0  \\
(0.20) $3$:& 0.9000         &     0   & 0.1000  \\
(0.14) $4$:&    0    &     0.8571   &      0.1429 \\ 
\bottomrule
\end{tabular}
\end{center}
}
Finally, consider the prior $p_3=(0.4,0.35,0.15,0.1)$,
which implies: $p_3(1)=0.4>1/k$, 
$p_3(2)=0.35>(0.35+0.15+0.1)/(k-1)=0.6/2=0.3$, and only 
$p_3(3)=0.15\leq(0.15+0.1)/(k-2)=0.25/1=0.25$. Therefore, $j^*=3$
and Step-2 of the algorithm becomes solving the following trivial 
system:
\begin{align*}
v_{\{1,2,3\}} = 0.15,& & &
v_{\{1,2,4\}} = 0.1,& & &
v_{\{1,2,3\}}, v_{\{1,2,4\}} \geq 0
\end{align*}
Hence, the corresponding optimal channel will be:
{\small\begin{center}
\begin{tabular}{@{} C{40pt} C{30pt} C{30pt} @{}}
\toprule
& $y_{\{\!1,2,3\!\}}$\quad  & $y_{\{\!1,2,4\!\}}$\quad  
\\
\cmidrule(r){2-3}
(0.40) $1$: & 0.6000  &  0.4000 \\
(0.35) $2$: & 0.6000  &  0.4000  \\ 
(0.15) $3$: &   1    &     0  \\ 
(0.10) $4$: &    0    &    1  \\ 
\bottomrule
\end{tabular}
\end{center}
}%
Note that the optimal channel 
always includes $1$ and $2$ in the pre-images, 
i.e., only shows $y_{\{\!1,2,3\!\}}$ and  $y_{\{\!1,2,4\!\}}$ outputs,
and, moreover, 
it induces $\pi=\left(p_3(1),p_3(2),p_3(3)+p_3(4)\right)=(0.4,0.35,0.25)$ posteriors for both of them.

We develop the proof of Theorem~\ref{Thm:main} in the following logical succession: 
First, we establish that $H(\pi)$ is an upper-bound for 
the posterior entropy $H(X|Y)$ for any feasible channel (Lemma~\ref{Lem:upper-bound}). 
Then we prove that this bound is tight by showing that Algorithm~\ref{Alg:Universal_Optimal_Cloaking} provides a feasible channel that achieves this upper-bound with equality, and hence,
is optimal (Lemma~\ref{Lem:strategy}).

\begin{lem}\label{Lem:upper-bound}
Given the prior $p$ and pre-image size cap $k$, for any feasible channel:  
$H(X|Y)\leq H(\pi)$.
\end{lem} 
\begin{lem}\label{Lem:strategy}
    For a given $p$ and $k$, Algorithm~\ref{Alg:Universal_Optimal_Cloaking} 
    produces a feasible channel 
    that achieves $H(X|Y)=H(\pi)$. 
\end{lem}

\begin{IEEEproof}[Proof of Lemma~\ref{Lem:upper-bound}]
Recall our generic form of conditional entropy in \eqref{eq:cond_entr_gen},  
where $\eta$ and $F$ satisfy 
\eqref{eq:subeq:g_inc_F_concave} or 
\eqref{eq:subeq:g_dec_F_convex}. 
We provide the proof for the case of 
\eqref{eq:subeq:g_inc_F_concave}. 
The treatment of case \eqref{eq:subeq:g_dec_F_convex} 
is similar. 
Since $F$ is symmetric and concave (case \eqref{eq:subeq:g_inc_F_concave}), it is also Schur-concave. 

Consider an arbitrary feasible channel satisfying the pre-image maximum size constraint. 
Then for 
any $y\in\mathcal{Y}^+$, we 
have $|\supp(p_{X|y})|\leq k$, 
that is, at most $k$ entries of $p_{X|y}$ 
are non-zero. This is due to the facts that $|\preimage(y)|\leq k$ 
and $p(x|y)=0$ for any $x\not\in\preimage(y)$.

Suppose that for the given $p$ and $k$, the value of  $j^*$ as defined in \eqref{eq:J_definition} is 1. For $j^*=1$, $\pi$
is the uniform distribution over $k$ elements, which is majorized by any probability distribution over a 
support size of at most $k$. Therefore, 
following Schur-concavity of $F$, each of the terms 
$F(p_{X|y})$  are bounded 
by $F(\pi)$. Hence, noting 
that $\eta$ is an increasing scalar function, we have:
\begin{multline*}
H(X|Y)=
\eta\left(\sum_{y\in\YY^+}
p(y)F(p_{X|y})\right)
\leq \eta\left(\sum_{y\in\YY^+}p(y)F\left(\pi\right)\right)\\= \eta\left(\!F\left(\pi\right)\!\sum_{y\in\YY+}\!p(y)\!\right)\!
=\eta\left(F\left(\pi\right)\right)= H(\pi)
\end{multline*}
This was intuitive: the highest uncertainty of the adversary, 
if the size of the pre-images is restricted to $k$, 
pertains to uniform distribution over the $k$ elements of the pre-image. 

Now, we turn our attention to cases where $j^*>1$.
First, $\forall y\in\YY^+$, following the 
symmetry of $F$, we can safely sort each of the posterior 
probabilities in non-increasing order, that is:
$F(p_{X|y})=F(p^{\downarrow}_{X|y})$. 

Second,  
the fact that $\forall y\in\YY^+$, $|\supp(p_{X|y})|\leq k$,
implies that the bottom $(n-k)$ elements of 
$p^{\downarrow}_{X|y}$ are always zero. 
Therefore, following the expansibility  of $F$, we can safely remove them. 
That is, $F(p^{\downarrow}_{X|y})=
F(p^{\downarrow}_{X|y\ \downarrow(1,\dotsc,k)})$, 
where the subscript $\downarrow\!(1,\dotsc,k)$ denotes 
projecting to only the first $k$ elements.

Third, note 
that $p(y)$ for $y\in\YY^+$ constitute coefficients 
of a convex combination, since each is non-negative and they add up to one. 
Hence, following the concavity property of $F$ (Jensen's 
inequality) and the 
previous two steps, we have:
\begin{multline}
H(X|Y)
=\eta\left(\sum_{y\in\YY^+}
p(y)F\left(p^{\downarrow}_{X|y\ \downarrow(1,\dotsc,k)}
\right)\right)\notag
\\\leq 
\eta\left(F\left(\sum_{y\in \YY^+}\!\!p(y)
p^{\downarrow}_{X|y\ \downarrow(1,\dotsc,k)}\right)\right)
\notag\\
=H\left(\sum_{y\in\YY^+}\!\!p(y)p^{\downarrow}
_{X|y\ \downarrow(1,\dotsc,k)}\right).
\label{eq:H_concavity_inequality}
\end{multline}   
The inequality in \eqref{eq:H_concavity_inequality} 
can be re-written as:
$H\left(X|Y\right)\leq H(q)$ where 
$q=(q_i), i=1,\dotsc,k$ is defined as follows:  
$q_i:=\sum_{y\in\YY^+}\left(p(y)p_{X|y}\right)_{[i]}$. Recall that subscript $[i]$ 
denotes the $i$'th largest element of a vector. 
Each vector $p(y)p_{X|y}$ is the joint probability distribution of $X$ and $Y$ for $Y=y$.
Specifically, we have:
\begin{gather*}
p(y)p_{X|y}=(p(y)p(x|y))_{x\in\XX} = (p(x,y))_{x\in\XX} = (p(x)p(y|x))_{x\in\XX}
\end{gather*}
Hence, we can rewrite  $q_i$ equivalently as $\sum_{y\in\YY^+}\left((p(x)p(y|x))_{x\in\XX}\right)_{[i]}$.
 
Fourth, we show that 
$q$, such defined, majorizes
$\pi$ as described in \eqref{eq:pi_j_definition}, i.e., $q\succ\pi$.
First of all,  $q$ is itself a probability distribution over a support of size $k$, 
since it is a convex combination of $k$-sized probability distributions 
$p^{\downarrow}_{X|y\ \downarrow(1,\dotsc,k)}$.
In particular, we have $\sum_{i=1}^kq_i=\sum_{i=1}^{k}\pi_i=1$.
Moreover, both $q$ and $\pi$ are already in non-increasing 
order: For $q$, this follows from the fact that all  
$p^{\downarrow}_{X|y\ \downarrow(1,\dotsc,k)}$ are in 
non-increasing order. For $\pi$, 
first note that its first $j^*-1$ entries match exactly those of the prior:  $(p(1),\ldots,p({j^*-1}))$,  
and are hence in non-increasing order according to our assumption for $p$. The next $k-j^*+1$ elements are all equal to $(\sum_{i=j^*}^np(i))/(k-j^*+1)$.
Hence, we just need to show 
$p(j^*-1)\geq (\sum_{i=j^*}^np(i))/(k-j^*+1)$.
This is a consequence of the definition of $j^*$. Specifically,  \eqref{eq:J_definition} implies that $p(j^*-1)>(\sum_{i=(j^*-1)}^np(i))/(k-(j^*-1)+1)$. Multiplying both side by $(k-(j^*-1)+1)$ and subtracting $p(j^*-1)$ from both sides yields our desired inequality.

Therefore, all we need to show in order to establish $q\succ\pi$ is that $\sum_{i=1}^lq_i\geq \sum_{i=1}^l\pi_i$ for all $l=1,\dotsc,(k-1)$. We will use the following sub-lemma: 
\begin{sub-lem}\label{sub-lem:Q_P}
$\sum_{i=1}^lq_i\geq \sum_{i=1}^lp(i)$ for any $l<k$. 
\end{sub-lem}	
\begin{IEEEproof}
Replacing for $q_i$, for any $l<k$, we have:
\begin{multline*}
\sum_{i=1}^lq_i=\sum_{i=1}^l\sum_{y\in\YY^+}\left((p(x)p(y|x))_{x\in\XX}\right)_{[i]}
\\=\sum_{y\in\YY^+}\sum_{\,\,i=1}^l\left((p(x)p(y|x))_{x\in\XX}\right)_{[i]}\geq 
\sum_{y\in\YY^+}\sum_{\,\,i=1}^lp(i)p(y|i)
\end{multline*}  
The second equality is simply switching the order of summations. The  inequality follows because summation of  the top $l$ elements of any vector is no less than the summation of any $l$ elements of it. The right hand side of the inequality, after a change in the order of summations, is equal to:
$\sum_{i=1}^l\sum_{y\in\YY^+}p(i)p(y|i)=\sum_{i=1}^lp(i)\sum_{y\in\YY^+}p(y|i)=\sum_{i=1}^lp(i)$. The last equality follows because  $\sum_{Y\in\YY^+}p(y|x)=1$ for each $x\in\XX$. 
Replacing this back in the inequality yields  $\sum_{i=1}^lq_i\geq \sum_{i=1}^lp(i)$, the claim of the  sub-lemma.
\end{IEEEproof}

Now, for any $l\leq j^*-1$, the inequality 
$\sum_{i=1}^lq_i\geq \sum_{i=1}^l\pi_i$ directly follows from the 
above sub-lemma, 
since  $\pi_i=p_i$ for all $i\leq j^*-1$ 
by its definition in \eqref{eq:pi_j_definition}. 
For an $l\in\{j^*,\dotsc, k-1\}$, first we argue 
that $\sum_{i=j^*}^lq_i/(l-j^*+1)\geq \sum_{i=j^*}^kq_i/(k-j^*+1)$: 
The left hand side is the (arithmetic) average of $(q_{j^*},\ldots,q_l)$, 
and the right hand side is the (arithmetic) average of $(q_{j^*},\ldots,q_k)$; 
the inequality then follows due to the fact that $q_i$'s are in 
non-increasing order.  This inequality can be written as 
$\sum_{i=j^*}^lq_i\geq\frac{l-j^*+1}{k-j^*+1} \sum_{i=j^*}^kq_i$. 
Adding $\sum_{i=1}^{j^*-1}q_i$ to both sides, and rewriting 
$\sum_{i=j^*}^kq_i$ equivalently as $(1-\sum_{i=1}^{j^*-1}q_i)$, 
we obtain:
$\sum_{i=1}^{l}q_i\geq \sum_{i=1}^{j^*-1}q_i +\frac{l-j^*+1}{k-j^*+1} 
(1-\sum_{i=1}^{j^*-1}q_i)$.
Following the sub-lemma, we have 
$\sum_{i=1}^{j^*-1}q_i\geq \sum_{i=1}^{j^*-1}p(i)$.
Now, consider the $\RR\to\RR$ function $f(x)=x+\frac{l-j^*+1}{k-j^*+1}(1-x)$. 
For any $j^*\in\{2,\ldots,k\}$, this function is increasing in $x$. 
Therefore, $\sum_{i=1}^{j^*-1}q_i\geq \sum_{i=1}^{j^*-1}p(i)$ implies 
$\sum_{i=1}^{j^*-1}q_i +\frac{l-j^*+1}{k-j^*+1} (1-\sum_{i=1}^{j^*-1}q_i)
\geq\sum_{i=1}^{j^*-1}p(i) +\frac{l-j^*+1}{k-j^*+1} (1-\sum_{i=1}^{j^*-1}p(i))$ 
as well. 
Note that the right hand side of the latter inequality is 
exactly $\sum_{i=1}^l\pi(i)$ when $l\in\{j^*-1,\ldots, k\}$.
Putting the cases of $l\leq j^*-1$ and $l\in\{j^*,\dotsc, k-1\}$ together, we obtain
$\sum_{i=1}^{l}q_i\geq\sum_{i=1}^l\pi_i$ for any 
$l\in{j^*,\ldots,k}$. 
This completes the argument for establishing $q\succ\pi$.

In the final step for proving Lemma~\ref{Lem:upper-bound}, we note that  Schur-concavity of $H$ together with  $q\succ\pi$ give $H(q)\leq H(\pi)$. The lemma now follows by noting that in step 3, we  showed 
$H(X|Y)\leq H(q)$.  
\end{IEEEproof}

Lemma~\ref{Lem:upper-bound} established that $H(\pi)$ is an upper-bound for the posterior entropy of any feasible channel. Next, we prove Lemma~\ref{Lem:strategy}, which states that our algorithm constructs a feasible channel that achieves this upper-bound, and hence, is optimal. Both lemmas hold for any symmetric expansible core-concave $H$.

\begin{IEEEproof}[Proof of Lemma~\ref{Lem:strategy}]
We provide the proof in the following sequence: 
(I): Algorithm~\ref{Alg:Universal_Optimal_Cloaking} indeed terminates with an output.
(II): The output of the algorithm is a feasible channel satisfying the pre-image size constraint. (III): The channel achieves $H(X|Y)=H(\pi)$.

(I): As we argued after 
\eqref{eq:J_definition}, $j^*$ can always be found.
Therefore, we only need to ensure that the linear system  
in Step 2 of the algorithm indeed has a solution. This is a consequence of the following sub-lemma:
\begin{sub-lem}\label{sub-lem:inside_feasible}
Consider a $(n-j+1)$-sized vector $p'=(p(j),\ldots,p(n))$ with non-negative elements, sorted in non-increasing order. Suppose $p(j)$, i.e., the biggest element of $p'$, satisfies 
$p(j)\leq {\sum_{i=j}^np(i)}/({k-j+1})$ for a $j$ and $k$,  $j\leq k\leq n$. 
Then the following system has a feasible solution:
\begin{align*}
\sum_{M\in\Env(\{1,\ldots,j-1,i\})} v_M= p(i), & & &\forall i=j,\ldots,n;\\
\text{subject to:}\ \ \ \ \ \ v_M \geq 0,& & & \forall M\in\Env(\{1,\ldots,j\!-\!1\}). 
\end{align*}
Moreover, for any solution we have:
\begin{gather*}
\sum_{M\in\Env(\{1,\ldots,j\!-\!1\})}v_M=\frac{\sum_{i=j}^{n}p(i)}{k-j+1}. 
\end{gather*}
\end{sub-lem} 

Note that the condition of  sub-lemma~\ref{sub-lem:inside_feasible} is satisfied for $j^*$ found in the first step of Algorithm~\ref{Alg:Universal_Optimal_Cloaking}.
\begin{IEEEproof}
For brevity, take $s:=\sum_{i=j}^np(i)$, $t:=(n-j+1)$, and $u:=(k-j+1)$.
Let:
\begin{gather*}
\Omega:=\{\omega\in{\mathbb{R}}^{t}:
\sum_{i=1}^t\omega_i= s,\ \&\ 0\leq \omega_i\leq \frac{s}{u}, \forall i=1,\dotsc,t\}. 
\end{gather*}
$\Omega$ is a \emph{convex polyhedron} in 
$\mathbb{R}^{t-1}$ (since it is described by a system of 
linear inequalities, and the minus 1 is due to the one equality 
constraint). It is also closed, and is non-empty, 
as $\omega=(s/t,\ldots,s/t)\in\Omega$. Hence, $\Omega$ is also a 
non-empty \emph{polytope} in $\mathbb{R}^{t-1}$, i.e., can 
be described as the convex hull of a finite number of points 
in $\mathbb{R}^{t-1}$. Specifically, any point inside $\Omega$ 
can be written as a convex combination of the \emph{extreme} (a.k.a. 
\emph{corner}) points of $\Omega$ (and {vice versa}). In fact, according 
to \emph{Carath\`{e}odory's theorem}, this can be done by a convex 
combination of at most $t$ of them. The extreme points of $\Omega$ are 
$t$-dimensional vectors where $w$ of their elements are $s/u$ and the $t-u$ 
rest of them are zeros. There are $\binom{t}{u}$ of such vectors. 
Let $\Lambda$ be a matrix whose columns are these extreme points, 
i.e, each column is a distinct permutations of $u$ entries of $s/u$ 
and $t-u$ entries of zero, that is: $\Lambda:=[(s/u,\ldots,s/u,0,\ldots,0)^T, 
\ldots,(0,\ldots,0,s/u,\ldots,s/u)^T]$.

The condition of the sub-lemma, i.e., $p(j)\leq {\sum_{i=j}^np(i)}/({k-j+1})$, or $p(j)\leq s/u$ implies that  $p'\in\Omega$. 
Hence, as we argued above, $p'$ can be expressed as a convex combination of 
the extreme points of $\Omega$. Let 
$z\in\mathbb{R}^{+{\binom{t}{u}}}$ denote such a convex combination, 
thus, we have:
$\Lambda z=p'$ where $z\geq 0$ 
(elementwise non-negative for all ${\binom{t}{u}}$ entries), and $\mathbf{1}^Tz =1$ where $\mathbf{1}$ is a ${\binom{t}{u}}$-sized vector of all ones. 

On the other hand, the linear system in the sub-lemma can be written in 
matrix form as: 
$\bar{\Lambda}v=p'$ where $\bar{\Lambda}$ is a $t\times \binom{t}{u}$ 
matrix whose columns are all the $\binom{t}{u}$ permutations of having 
$u$ entries of 1 and $t-u$ entries of 0. Therefore, 
with some re-ordering of the equations if necessary, we can write: 
$\Lambda=(s/u)\bar{\Lambda}$. Hence, $\Lambda z=p'$ implies 
$(s/u)\bar{\Lambda} z=p'$, and $z\geq 0$ implies 
$(s/u)z\geq 0$. Therefore, $v=(s/u) z$ is a 
feasible solution of the system in the sub-lemma. 

The second claim of the sub-lemma follows from summating all the 
equations of the system and a simple counting:
$\sum_{i=j}^n p(i)$ $=\sum_{i=j}^n	\sum_{M\in\Env(\{1,\ldots,j-1,i\})} v_M$ 
$=  (k-j+1)\sum_{M\in\Env(\{1,\ldots,j\!-\!1\})}v_M$.
\end{IEEEproof}
This finishes part (I) of the lemma's proof: that 
Algorithm~\ref{Alg:Universal_Optimal_Cloaking} always terminates with a solution.

(II): 
First, note that the algorithm  assigns a non-zero value to $p(y_M|i)$ only for $i\in M$. Hence, the pre-image of $y_M$ is a subset of $M$, and thus, its size is bounded by the size of $M$, i.e., $k$. 
Specifically, for an $i\in\{j^*,\dotsc, n\}$, Algorithm~\ref{Alg:Universal_Optimal_Cloaking} 
assigns $p(y_M|i)= v_M/p(i)$ for all $M\in\Env(\{1,\ldots,j^*\!-\!1,i\})$, 
and zero for any other $y$. 
Hence, $\sum_{y\in\YY}p(y|i)
= \sum_{M\in\Env(\{1,\ldots,j^*\!-\!1,i\})}p(y_M|i)=\sum_{M\in\Env(\{1,\ldots,j^*\!-\!1,i\})}v_M/p(i)=1$, 
where the last equality follows directly from the system 
of equations in Step~2 of the algorithm, specifically, the 
equality constraint of $\sum_{M\in\Env(\{1,\ldots,j^*\!-\!1,i\})}v_M=p(i)$.
Similarly, for an $i\in\{1,\ldots,j^*-1\}$, 
the algorithm assigns: $p(y_M|i)=v_M(k-j^*+1)/\sum_{j=j^*}^np(j)$ 
for all $M\in\Env(\{1,\ldots,j^*\!-\!1\})$ and zero for any 
other $y$. Therefore, $\sum_{y\in\YY}p(y|i) = \sum_{M\in\Env(\{1,\ldots,j^*\!-\!1\})}p(y_M|i)
=\sum_{M\in\Env(\{1,\ldots,j^*\!-\!1\})}v_M(k-j^*+1)/\sum_{j=j^*}^np(j)=1$,
where the last equality is due to the second claim of
Sub-lemma~\ref{sub-lem:inside_feasible}, 
that $\sum_{M\in\Env(\{1,\ldots,j^*\!-\!1\})}v_M=
(\sum_{i=j^*}^{n}p(i))/(k-j^*+1)$.
Hence, Algorithm~\ref{Alg:Universal_Optimal_Cloaking} terminates with 
a valid channel that satisfies the pre-image size constraints.

(III): 
The pre-images of the channel constructed by the algorithm are $M\in\Env(\{1,\dotsc,j^*\!-\!1\})$ for which $v_M>0$. In particular,
all of these pre-images include inputs $1,\dotsc,{j^*-1}$, along with $k-j^*+1$ other inputs. 
Let $M=\{1,\ldots, j^*-1,\phi_1,\ldots,\phi_{k-j^*+1}\}$, 
where $\{\phi_1,\dotsc,\phi_{k-j^*+1}\}\subset \{j^*,\dotsc,n\}$ 
be any of such pre-images for which $v_M>0$. The posterior probability 
distribution for $y_M$ 
is given by the Bayes' 
rule: $p(x|y_M)={p(x)p(y_M|x)}/p(y_M)$ 
where $p(y_M)=\left(\sum_{x'\in \XX}p(x')p(y_M|x')\right)$.
Replacing from the assignments in Steps~3 through 5 of 
Algorithm~\ref{Alg:Universal_Optimal_Cloaking}, we get:
\begin{multline*}
p(y_M)=\sum_{i=1}^{j^*-1}p(i)\left(\frac{v_M(k-j^*+1)}{\sum_{j=j^*}^np(j)}\right)+\sum_{i=1}^{k-j^*+1}p(\phi_i) \frac{v_M}{p(\phi_i)}\\
=v_M(k-j^*+1)\left(\frac{\sum_{i=1}^{j^*-1}p(i)}{\sum_{j=j^*}^np(j)}+1\right)=
\frac{v_M(k-j^*+1)}{\sum_{j=j^*}^np(j)}
\end{multline*}
Hence, for all $i=1,\dotsc,k-j^*+1$:
\begin{gather}\label{eq:Algo_inducing_first}
p(\phi_i|y_M)=
\frac{p(\phi_i) v_M/p(\phi_i)}{v_M(k-j^*+1)/
\sum_{j=j^*}^np(j)}=\frac{\sum_{j=j^*}^np(j)}{k-j^*+1}
\end{gather}
On the other hand, for $i=1,\dotsc,j^*-1$:
\begin{gather}\label{eq:Algo_inducing_second}
p(i|y_M)=\frac{p(i)v_M(k-j^*+1)/
\sum_{j=j^*}^np(j)}{v_M(k-j^*+1)/\sum_{j=j^*}^np(j)}=p(i)
\end{gather}
According to \eqref{eq:Algo_inducing_first} and \eqref{eq:Algo_inducing_second} and the definition of $\pi$ in \eqref{eq:pi_j_definition},
a channel resulting from 
Algorithm~\ref{Alg:Universal_Optimal_Cloaking} ensures that
for each $y\in\YY^+$,  $p_{X|y}=\pi$.
Therefore, employing such a channel, we will have:
\begin{multline*}
H(X|Y)=
\eta\left(\sum_{y\in\YY^+}
p(y)F\left(p_{X|y}\right)\right)\\ 
=\eta\left(\sum_{y\in\YY^+}p(y)F\left(\pi\right)\!\right)\!
=\eta\left(F\left(\pi\right)\right)= H(\pi).
\end{multline*}
This concludes the proof of 
Lemma~\ref{Lem:strategy}, and thus, of  Theorem~\ref{Thm:main}. 
\end{IEEEproof}

In what follows, in order to showcase the versatility of our main result, we provide a series of corollaries.
\begin{corr}\label{prop:opt_min_ent}
    Given prior $p$ and pre-image size-cap $k$, the maximum achievable posterior entropy with respect to  
    Min-Entropy is $-\log(\max(1/k,p_{[1]}))$.  This in turn implies that the minimum achievable leakage with respect to Min-Entropy is  
    $0$ for any $k\geq 1/p_{[1]}$, and  $-\log(kp_{[1]})$ for $k< 1/p_{[1]}$.
\end{corr}
\begin{IEEEproof}
    From Theorem~\ref{Thm:main}, 
    if $p_{[1]}\leq 1/k$, then $j^*=1$ and $\pi=(1/k,\ldots,1/k)$, which means the highest achievable posterior entropy 
    is $H((1/k,\ldots,1/k))$. For Min-Entropy, this gives  $-\log(1/k)$. 
    If on the other hand $p_{[1]}>1/k$, then $j^*$ is an index between $2$ and $k$. 
    For any $j^*>1$, the largest element of $\pi$ is $p_{[1]}$, hence $H(\pi)$ 
    for Min-Entropy is equal to $-\log(p_{[1]})$. Putting these together yields the claim. 
\end{IEEEproof}

Corollary~\ref{prop:opt_min_ent} may come as a bit of a surprise:  if $p_{[1]}>1/k$, 
the information leakage with respect to Min-Entropy can be made absolutely zero. 
This can in fact be generalized to $l$-Guess-Entropy too: 			 
If $p_{[l]}\geq \left({\sum_{i=l}^kp_{[i]}}\right)\Big/(k-l+1)$, then the minimum 
leakage with respect to $l$-Guess-Entropy is zero. These results however do not 
contradict the Shannon's perfect secrecy, since, unlike Shannon's entropy,
Min-Entropy and $l$-Guess-Entropy do not retain the information of the whole distribution.
Also note that these zero-leakage cases correspond to priors that are highly skewed. 
In such cases, the prior is already very revealing and gives a big advantage to the adversary, but the defender can 
at least leverage those high probability inputs to not reveal any extra information.  In other words, figuratively speaking, the inputs with high probabilities cannot be helped, but the small-probability inputs can ``hide in their shadow''. 

Extension of Corollary~\ref{prop:opt_min_ent} to the $l$-Guess and Guesswork are provided next (proofs skipped for brevity). 
\begin{corr}
   Given a prior $p$ and pre-image size-cap $k$, the maximum  posterior entropy with respect to  
    $l$-Guess-Error-Probability, i.e., the probability that an adversary is wrong within his $l$ best guesses,  is:
    \begin{gather*} 
    1-\max_{0\leq j\leq l} \bigg\{ \sum_{i=1}^{j}p_{[i]}+\big(1-\sum_{i=1}^jp_{[i]}\big)\frac{l-j}{k-j}\bigg\}
    \end{gather*}
    \end{corr}
    \begin{corr}
    Given a prior $p$ and pre-image size-cap $k$, the maximum  posterior entropy with respect to the Guesswork entropy, i.e., the expected number of guesses of an adversary before detection, is: 
        \begin{gather*}
        \min_{1\leq j\leq k} \bigg\{ \sum_{i=1}^{j-1}ip_{[i]}+\big(1-\sum_{i=1}^{j-1}p_{[i]}\big)\frac{k+j}{2}\bigg\}
        \end{gather*}
    \end{corr}
 
\subsection{A counterexample to metric invariance}
 It is not possible in general to have an optimal solution that is metric invariant.
 The following is a counterexample: consider  $4$ secrets: $\{1,2,3,4\}$ with prior $(p_1,p_2,p_3,p_4)$. The set of observables (outputs) is $\{a,b\}$. The set of feasible observables is defined by $\Omega=\{(1,a),(2,a),(2,b),(3,b),(4,b)\}$. That is, for secret $1$, the only possible observable to show is $a$, for secret $2$, both  $a$ and $b$ are allowed, and for secrets $3$ and $4$, the only allowed observable is $b$. Following the admissible observables for secrets $1$, $3$ and $4$, we have: $\delta(b|1)=\delta(a|3)=\delta(a|4)=0$, and therefore:
  $\delta(a|1)=\delta(b|3)=\delta(b|4)=1$. For secret 2, $\delta(a|2)$ and $\delta(b|2)$ are free, as long as they are positive and add up to $1$. Therefore, $\delta(b|2)$ is the only variable of optimization. 
 For this example the optimal depends on the measure chosen:
setting $\delta(b|2)=x/p_2$
we have that the maximizer $x$ for guesswork entropy is 0.1518, for  R\'enyi with $\alpha=2$ is  0.2573, for Shannon entropy is 0.2998.

\section{Departure From Symmetric Entropies: Extension to
	Gain-Based Leakages}\label{sec:Gains}

In the previous section, we provided a  channel
that,  under a pre-image size constraint, yields
minimum leakage with respect to a large class of classical entropy measures.
Our analysis only relied on structural properties of the entropy function,
namely: symmetry, expansibility, and   core-concavity.
A major point of departure from this family of entropies, where potentially
all three of these properties can be violated, is the gain based entropy
($g$-entropy)
introduced in \cite{alvim2012measuring}. $g$-entropy is a generalization of the notion of
Min-Entropy  by permitting secret-guess dependent gains to a guessing adversary.
This notion of leakage has received attraction in a line of research
(e.g. \cite{kopf2013preventing,biondi2013quantifying,mciver2014abstract,mardziel2014quantifying,alvim2014additive}).
We now introduce our  generalization of $g$-entropy by fusing it with a generic classical entropy,
and present an extension of our main result.

Given a set of guesses $\mathcal{W}$ and secrets $\XX$, we start by defining a gain function $g$ by a matrix $G\in \mathbb{R}^{|\mathcal{W}|\times
	|\XX|}$, where $G_{w,x}:=g(w,x)$. The coefficient $g(w,x)$ is the gain of the adversary when her guess is $w$ and the secret is $x$. We consider now a gain matrix $G$ such that the vector $Gp/\|Gp\|_1$ is elementwise non-negative for any probability distribution $p$ over a fixed support (and hence, $Gp/\|Gp\|_1$ is a legitimate probability distribution). Then a generalized gain-based entropy and its corresponding conditional entropy can be defined as follows:
\begin{gather}\label{eq:def_entropy_gain_general}
H_{g}(X):=\eta\left(\|Gp\|_1 F\left(\frac{Gp}{\|Gp\|_1}\right)\right)
\\ H_g(X|Y)\!:=\!
\eta\left(\sum_{y\in\YY^+}
p(y)\|Gp_{X|y}\|_1F\left(
\frac{Gp_{X|y}}{\|Gp_{X|y}\|_1}\!\right)\!\!\right)
\label{eq:def_conditional_entropy_gain_general}
\end{gather}
where, as before, $\eta$ is a monotonic scalar function and $F$ is a symmetric expansible core-concave function.
For instance, 
$g$-entropy  \cite{alvim2012measuring}
is retrieved by taking $\eta(\cdot)=-\log(\cdot)$ and
$F(\cdot)=\|\cdot\|_{\infty}$.
In fact, a whole family of entropies can be derived from the R\'enyi family, $H_\alpha$, by
taking $\eta(\cdot)=\frac{-\alpha}{\alpha-1}\log(\cdot)$ and
$F(\cdot)=\|\cdot\|_{\alpha}$
(noting the scalability of the $\alpha$-norm) as follows:
$H_{\alpha,g}(X):=H_{\alpha}(Gp)=
\frac{-\alpha}{\alpha-1}\log\|Gp\|_{\alpha}$.
All R\'enyi entropies are trivially instances
of this $(\alpha,g)$ family by taking $G$ to be the identity matrix.
In particular, Shannon entropy is retrieved by also letting $\alpha\to 1$.

The $g$-leakage defined in
\cite{alvim2012measuring} can now be
generalized as the difference between prior
and posterior entropies defined in \eqref{eq:def_entropy_gain_general}
and \eqref{eq:def_conditional_entropy_gain_general} respectively.
In what follows, we establish that the leakage such defined
is always non-negative (hence generalizing
\cite[Theorem 4.1]{alvim2012measuring}).  The proof is similar to that of Proposition~\ref{prop:leakage_positive_dpi} and is removed for brevity.

\begin{prop}\label{prop:Leakage_positive_gen}
	$H_g(X) - H_g(X|Y)\geq 0$.
\end{prop}

Note that for almost any $G$ other than the identity matrix, our new entropy
functions $H_g$
are no longer symmetric (nor expansible or core-concave)
in $p$.  However, for a special class of matrix gains, namely
\emph{diagonal} matrices, we present a generalization of Theorem~\ref{Thm:main}.
We use the notation $G=\diag(\gamma)$ where
$\gamma=(\gamma_1,\dotsc,\gamma_n)\in \mathbb{R}^{+n}$, to
indicate that $G$
is a square diagonal matrix (zero for every entry except for
possibly the diagonal
elements). This models cases where the adversary gains $\gamma_i\geq 0$
if the channel's input is $i$ and he identifies it correctly, and zero if
he mis-identifies.
Although investigating only diagonal gain matrices may be restrictive,
they do exhibit the secret-dependent non-symmetric
essence of the $g$-leakage.

\begin{prop}\label{Thm:gain-extension}
	Let $p=(p(1),\dotsc,p(n))$ be the prior,   $G=\diag(\gamma)$, be
	the diagonal gain matrix with non-negative gains
	(with at least one of them strictly positive),
	and let $k$ be the maximum allowed size of the pre-images.
	Without loss of generality, assume that $Gp=(\gamma_1p(1),\dotsc,
	\gamma_np(n))$ is
	in non-increasing order.
	Then Theorem~\ref{Thm:main} holds with $p$ replaced with $Gp$. Specifically, let index $j^*$ and vector $\pi=(\pi_1,\ldots,\pi_k)\in\mathbb{R}^{+k}$ be defined as follows:
	\begin{equation}\label{eq:definition_pi_generalized}
	\begin{aligned}
	j^*:=\min\left\{j: 1\leq j\leq k, \gamma_jp(j)\leq \dfrac{\sum_{i=j}^{n}\gamma_ip(i)}{k-j+1}\right\},\\
	\!\pi_l\!:=\!\gamma_lp(l):l\leq j^*\!-\!1,\ 
	\pi_l\!=\!\frac{\sum_{i=j^*}^n\gamma_ip(i)}{k-j^*+1}:j^*\!\leq\! l\!\leq\! k.
	\end{aligned}
	\end{equation}
	Then:
	\begin{itemize}
		\item[A.] The maximum achievable posterior entropy
		$H_g(X|Y)$ among all feasible channels satisfying pre-image size constraint is
		$\eta\left(\|\pi\|_1F\left({\pi}/{\|\pi\|_1}\right)\right)$.
		\item[B.] Algorithm~\ref{Alg:Universal_Optimal_Cloaking} where $Gp$ (ordered in decreasing order) replaces $p$ (and assuming the convention of $0/0=0$ whenever necessary) explicitly provides a feasible (randomized) channel that achieves
		the above maximum posterior entropy for any
		generalized measure per \eqref{eq:def_conditional_entropy_gain_general}.
	\end{itemize}
\end{prop}

The extension makes intuitive sense: The gain coefficients, $\gamma_i$'s,
represent the relative importance of having a secret revealed.
The algorithm multiplies each probability by its corresponding gain and
tries to make this effective importance of the secrets as uniform as
possible.
The proof is very similar
to that of Theorem~\ref{Thm:main}. We hence omit the detail and just provide an overview of it.
As before, one can first establish that
$\eta\left(\|\pi\|_1F\left({\pi}/{\|\pi\|_1}\right)\right)$ is an upper-bound for
$H_g(X|Y)$ for any feasible channel, and
then  prove that 
Algorithm~\ref{Alg:Universal_Optimal_Cloaking}, fed with $Gp$ instead of $p$, produces a valid
channel that achieves this upper-bound with equality and is hence optimal. Notably, the arguments again hold for any choice of the entropy in this family (for a fixed gain matrix), and therefore, the optimality of the provided channel is, once again, metric-invariant.

\section{Numerical Illustrations}\label{sec:Numerics}
First, we investigate the effect of the maximum permitted pre-image size, $k$, and the choice of the entropy on the minimum achievable leakage. 
We consider three candidate entropies: Shannon, Guesswork, and Min-Entropy. 
Recall that: $H_{\mathrm{Sh.}}(X):=-\sum_{x\in\XX}^np(x)\log(p(x))$ and its 
posterior entropy is $H_{\mathrm{Sh.}}(X|Y)=\sum_{y\in\YY^+}p(y)\left(-\sum_{x\in\XX}^np(x|y)\log(p(x|y))\right)$. 
For Min-Entropy,  $H_{\infty}(X):=-\log \max_{x\in\XX}(p(x))$, and posterior entropy is computed as $H_{\infty}(X|Y)=-\log(\sum_{y\in\YY^+}p(y)\max_{x\in\XX}(p(x|y))$,  a case of \eqref{eq:subeq:g_dec_F_convex}.  For Guesswork, $H_{\mathrm{Gu.}}(X)=\sum_{i}^nip_{[i]}$ and the posterior entropy is $H_{\mathrm{Gu.}}(X|Y)=\sum_{y\in\YY^+}\sum_{i}^ni(p_{X|y})_{[i]}$,  a case of \eqref{eq:subeq:g_inc_F_concave}. To obtain a comparable scale for all three, we added a $\log(\cdot)$ to both prior and posterior of Guesswork entropy as well.  

For all examples in this section, we consider a input space consisting of $30$ elements with the following prior distribution: $p=(30/465,29/465,\dotsc,1/465)$.
Fig.~\ref{Fig:min_leakage_vs_k} shows that, as we expect, the minimum leakage reduces as larger pre-images are allowed. When leakage is quantified with Shannon entropy, min-leakage only vanishes when $k=n$, in accordance with the classic perfect secrecy result. However, the minimum achievable information leakage with respect  to Min-Entropy becomes zero for any $k\geq 16$ in our example. This complies with the result of Corollary~\ref{prop:opt_min_ent}, which stated that for any $k\geq \lceil 1/p_{[1]} \rceil$, an optimal channel can achieve zero leakage with respect to Min-Entropy. In this example, $\lceil1/p_{[1]}\rceil=\lceil465/30\rceil=16$.      

\begin{figure*}[ht]
    \centering
    \begin{subfloat}[][]{\label{Fig:min_leakage_vs_k}
            \includegraphics[width=0.32\textwidth]{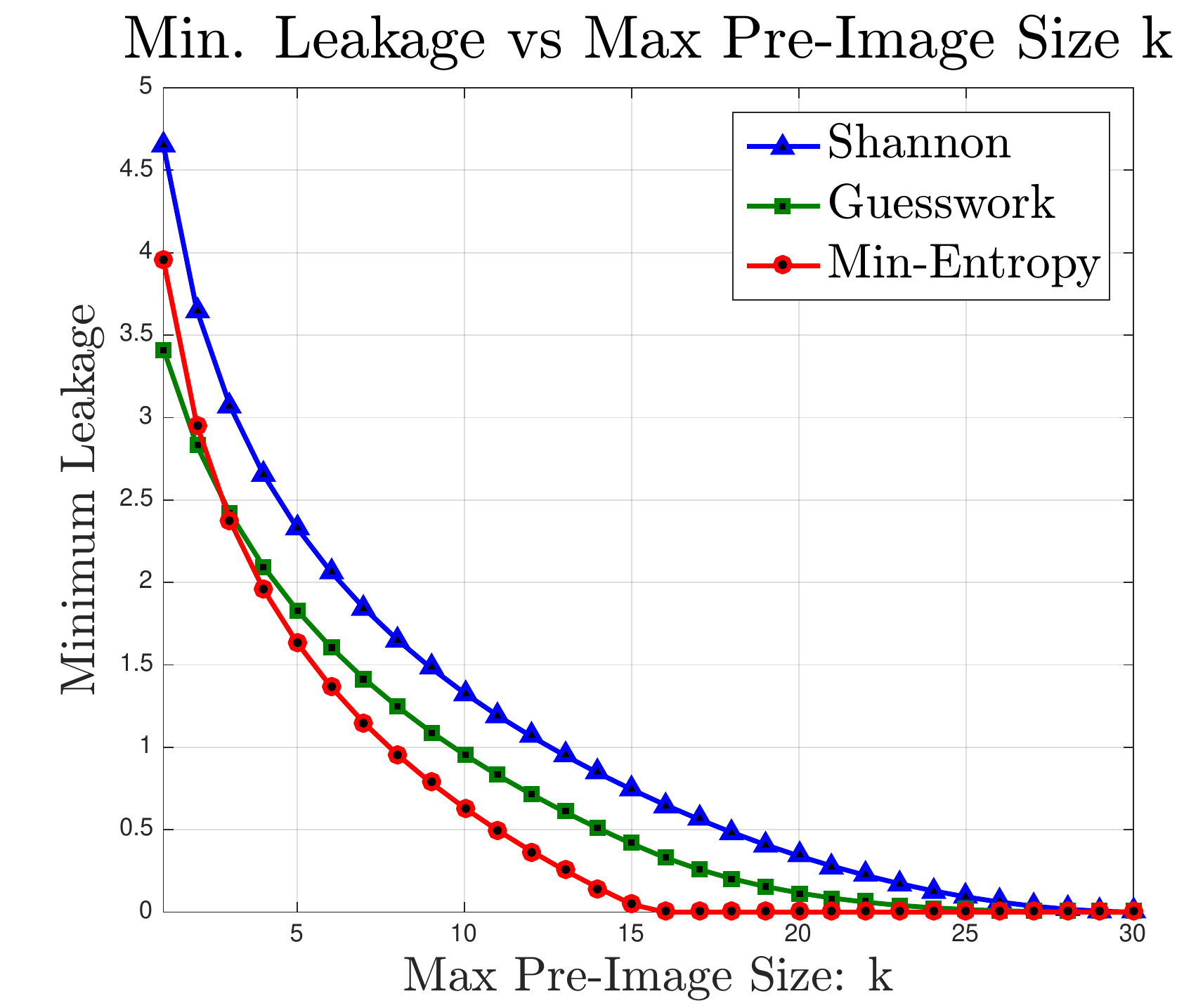}}
    \end{subfloat}\ 
    \begin{subfloat}[][]{\label{Fig:opt_vs_unif_varying_k}
            \includegraphics[width=0.32\textwidth]{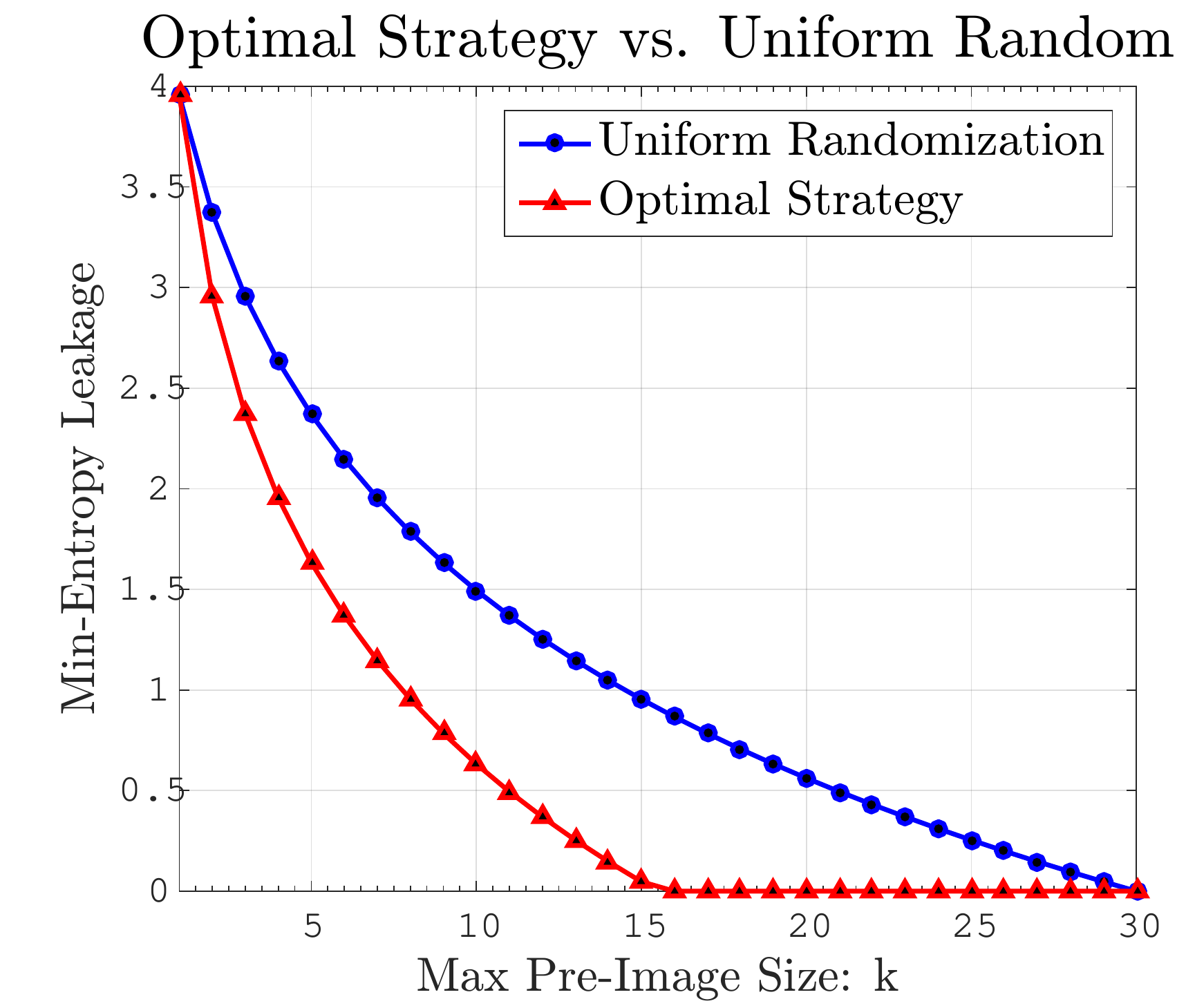}}
    \end{subfloat}\ 
    \begin{subfloat}[][]{\label{Fig:informed_vs_ignorant_varying_k}
            \includegraphics[width=0.32\textwidth]{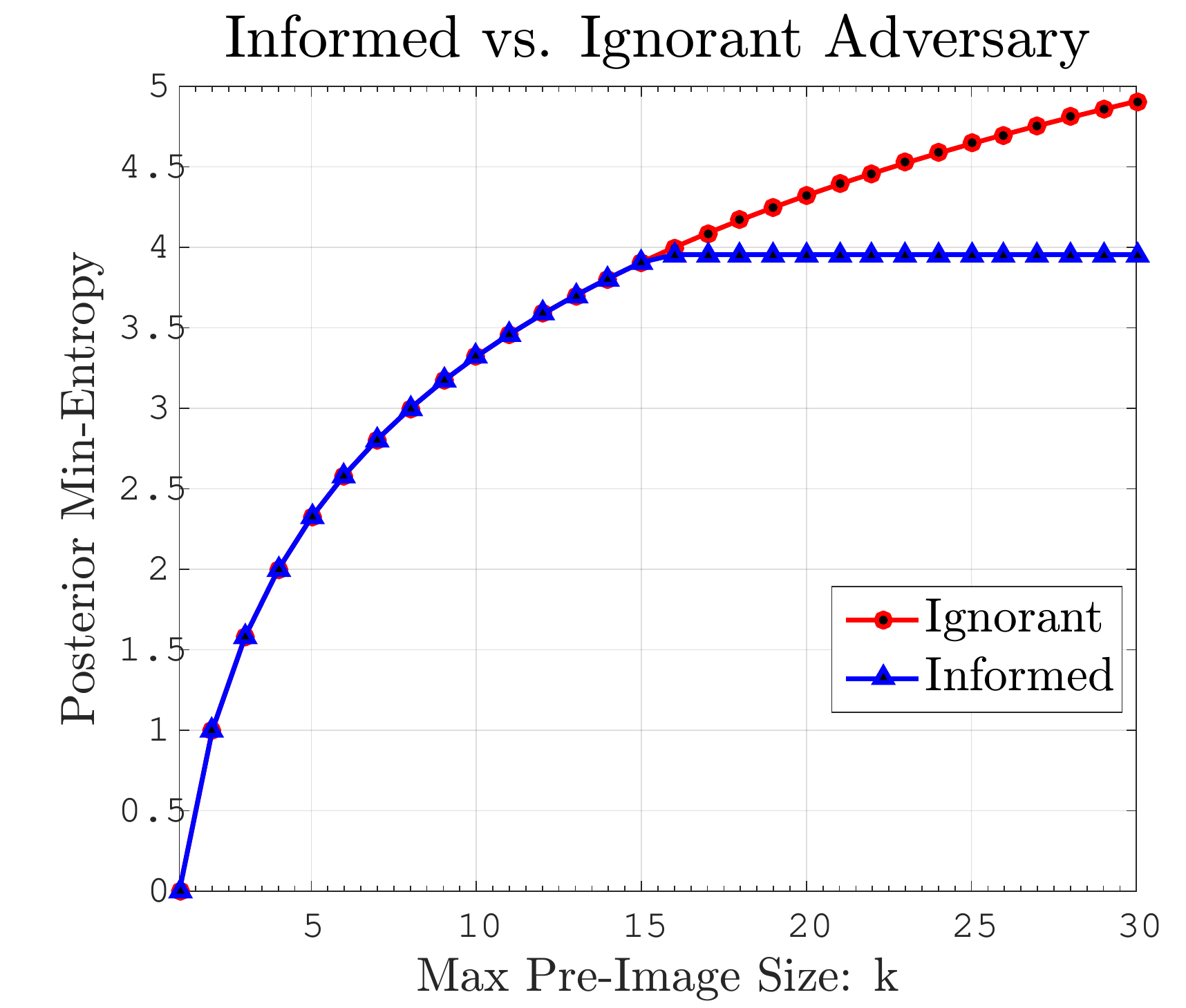}}
    \end{subfloat}
    \caption{For all figures, the prior is $p=(30/465,\ldots,1/465)$ and the pre-image size-cap, $k$, is varied from $1$ to $n=30$ as the x-axis.  (a)~The minimum achievable leakage  with respect to Shannon, Guesswork and Min-Entropy. The minimum leakage improves as larger pre-images are allowed. The Shannon entropy only becomes zero for $k=n$, as is the classic perfect secrecy, while the best min-entropy leakage becomes zero  for any $k\geq 16$ for this prior distribution as per Corollary~\ref{prop:opt_min_ent}. (b)~Comparison of the Min-Entropy leakage achieved by the optimal channel and the base-line (uniform randomization) strategy. (c)~Negative of the log of the expected reward of an informed adversary, who knows the true prior distribution of the channel input, and an uninformed  adversary who simply assumes a uniform prior. The channel (randomized strategies) is designed to be optimal assuming facing an informed adversary.}
\end{figure*}

Next, we compare the performance of optimal channels against the following base-line: For a given input, construct its set of maximal pre-images to be the subsets of size $k$ of the inputs that include that particular input, i.e., is composed of that input and $k-1$ others. Then \emph{uniformly randomly} pick the outputs that correspond to those pre-images. Note that this strategy is in fact optimal when the prior distribution is uniform, but not necessarily for other priors.   Fig.~\ref{Fig:opt_vs_unif_varying_k} depicts the leakage with respect to Min-Entropy achieved by the optimal strategy and the base-line strategy when $p=(30/465,\ldots,1/465)$, demonstrating the sub-optimality of the base-line for any intermediate value of $k$.  Adoption of this strategy is sub-optimal because it essentially ignores the fact that an adversary who is aware of the distribution of the input can exploit it to further improve his guessing power.

Next, we investigate the effect of one of the assumptions we made in the paper: that the defender designs her channel assuming that the adversary knows the true distribution of the inputs. In particular, we consider an uninformed adversary, that does not know the prior distribution, and thus, for any observed output, simply chooses a guess uniformly randomly. What will be the performance of the strategy that is designed to be optimal with the (worst-case) assumption that the adversary is informed of the true distribution, but facing an uninformed (ignorant) adversary instead.  
In Fig.~\ref{Fig:informed_vs_ignorant_varying_k}, for the prior of $p=(30/465,\ldots,1/465)$, we have depicted the posterior Min-Entropy for an informed vs. ignorant adversary. As we can see,  for $k\geq 16$, the Min-Entropy of the ignorant adversary is larger than that of the informed one. For $k<16$, we have $p_{[1]}<1/k$, which implies $j^*=1$, and hence, the  optimal channel indeed induces uniform posterior distributions on the pre-image of any shown output. Hence, uniformly random guessing from any observed output by the uninformed adversary matches the optimal strategy of an informed adversary as well.

\section{Conclusion and Future Work}
We investigated the problem of minimizing leakage when perfect 
secrecy is not achievable due to operational limits on the allowable size of 
the conflating sets.  
We constructively shown the existence of metric-invariant optimal channels 
achieving minimum leakage for any choice of entropy that satisfy a 
mild set of conditions (symmetry, expansibility, and
core-concavity).

We expect that the techniques developed in our proofs, 
especially majorization arguments, be reused
in unification of different notions of leakage and 
establishing robustness results for more general set of constraints. Exploring concrete application-oriented settings, e.g., in side-channel defence and attack, is another goal in our future research. Extensions of our framework to leakage metrics that consider worst-case scenarios like maximal leakage \cite{issa2016operational} and differential privacy are also yet to be investigated.

\bibliographystyle{IEEEtran}

\begin{IEEEbiography}[{\includegraphics[width=1in,height=1.25in,clip,keepaspectratio]{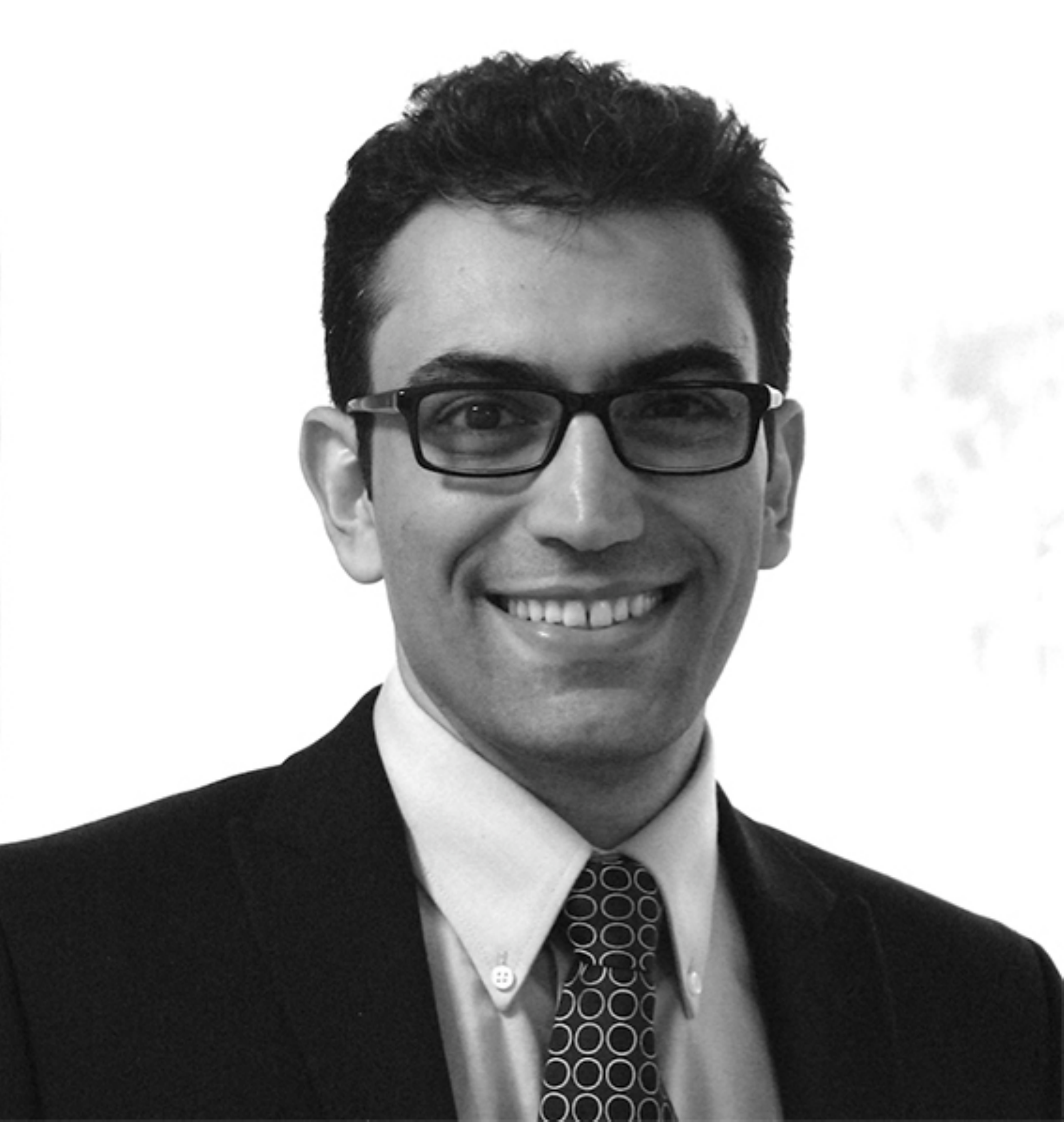}}]{MHR Khouzani}
received his Ph.D. in Electrical and Systems Engineering in 2011 from University of Pennsylvania. He held postdoctoral research positions with the Ohio State University (OSU), the University of Southern California (USC), Royal Holloway, University of London (RHUL), and Queen Mary, University of London (QMUL). Since November of 2016, he is a Lecturer  in the EECS department at QMUL.
Dr. Khouzani's research is in the area of  information security. He uses analytical tools from areas such as information theory, optimization, and game theory, to contribute to field of the science of security.
\end{IEEEbiography}

\begin{IEEEbiography}[{\includegraphics[width=1in,height=1.25in,clip,keepaspectratio]{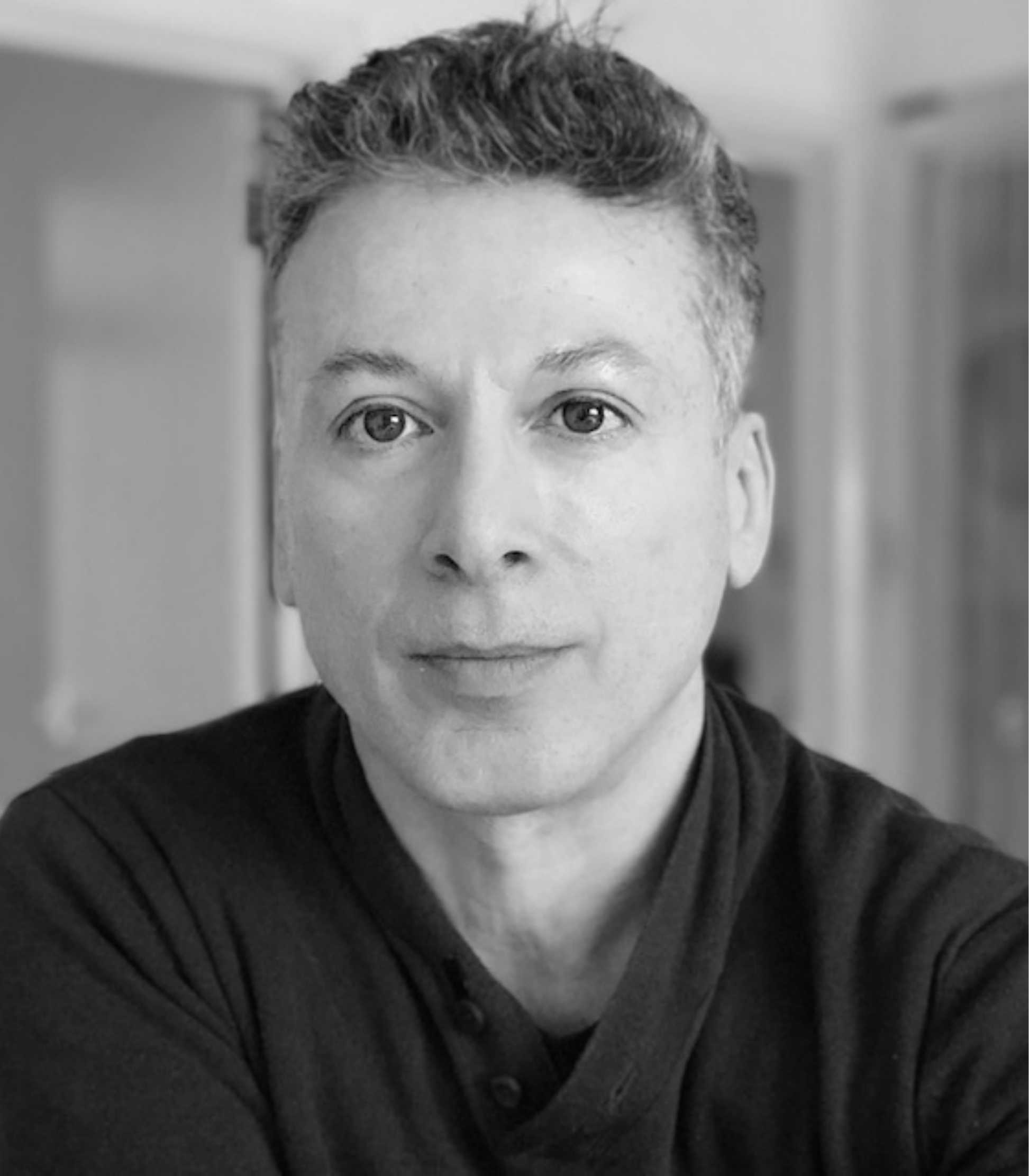}}]{Pasquale Malacaria}
 received his Laurea in Philosophy from "La Sapienza"
	University in Rome and his PhD in ``Logique et fondements de
	l'Informatique'' from the University of Paris VII in France. His work
	focuses on information theory, game theory, verification and their
	applications to computer security.
	He is a Professor of Computer Science at Queen Mary University of
	London. He has been an EPSRC advanced research fellow, is a recipient of
	the Alonzo Church award 2017 and of the Facebook Faculty awards 2015. 
\end{IEEEbiography}

\end{document}